\RequirePackage{ifpdf}
\ifpdf 
\documentclass[pdftex]{sigma}
\else
\documentclass{sigma}
\fi

\usepackage{multirow}
\usepackage{slashed}

\def\obar{\overline}

\newcommand{\barr}{\begin{array}}
\newcommand{\earr}{\end{array}}

\def\a{\alpha}  \def\b{\beta}
 \def\g{\gamma} \def\G{\Gamma}
 \def\d{\delta} \def\D{\Delta}
 \def\e{\epsilon} 
\def\f{\phi} \def\F{\Phi}   
\def\l{\lambda}  \def\la{\lambda} 
 \def\o{\omega}   
\def\s{\sigma}  \def\t{\tau}

  \def\cC{{\cal C}}    
\def\cH{{\cal H}}      \def\cN{{\cal N}}
    \def\cS{{\cal S}}

\def\R{{\mathbb R}} \def\C{{\mathbb C}} 
\def\Z{{\mathbb Z}} \def\one{\mbox{1 \kern-.59em {\rm l}}}

\def\bit{\begin{itemize}} \def\eit{\end{itemize}} \def\Tr{\mbox{Tr}}

\def\({\left(} \def\){\right)} \def\tens{\otimes}

\numberwithin{equation}{section}

\begin{document}

\newcommand{\miso}{\frac{1}{2}}

\def\mc{\mathcal}

\newcommand{\fet}{\frac{1}{3}}
\newcommand{\fdt}{\frac{2}{3}}
\newcommand{\ftt}{\frac{4}{3}}
\def\w{\wedge}
\def\olra{\overleftrightarrow}
\def\vf{\varphi}
\def\g{\gamma}
\def\1{\frac{G_2}{SU(3)}}
\def\2{\frac{Sp_4}{SU(2)\times U(1)}}
\def\3{\frac{SU(3)}{U(1)\times U(1)}}
\def\t{\tilde}
\def\l{\lambda}
\def\a{\alpha}

\allowdisplaybreaks

\renewcommand{\thefootnote}{$\star$}

\renewcommand{\PaperNumber}{063}

\FirstPageHeading

\ShortArticleName{Higher-Dimensional Unif\/ied Theories with Fuzzy Extra Dimensions}

\ArticleName{Higher-Dimensional Unif\/ied Theories\\ with Fuzzy Extra Dimensions\footnote{This paper is a
contribution to the Special Issue ``Noncommutative Spaces and Fields''. The
full collection is available at
\href{http://www.emis.de/journals/SIGMA/noncommutative.html}{http://www.emis.de/journals/SIGMA/noncommutative.html}}}

\Author{Athanasios CHATZISTAVRAKIDIS~$^{\dag \ddag}$ and George ZOUPANOS~$^\ddag$}

\AuthorNameForHeading{A. Chatzistavrakidis and G. Zoupanos}

\Address{$^\dag$~Institute of Nuclear Physics,
NCSR  Demokritos,
GR-15310 Athens, Greece}
\EmailD{\href{mailto:cthan@mail.ntua.gr}{cthan@mail.ntua.gr}}

\Address{$^\ddag$~Physics Department, National Technical University of Athens,\\
\hphantom{$^\ddag$}~GR-15780 Zografou Campus, Athens, Greece}
\EmailD{\href{mailto:george.zoupanos@cern.ch}{george.zoupanos@cern.ch}}

\ArticleDates{Received May 06, 2010, in f\/inal form July 22, 2010;  Published online August 12, 2010}

\Abstract{Theories def\/ined in higher than four dimensions have been used in various frameworks and have a long and interesting history. Here we review certain attempts, developed over the last years, towards the construction of unif\/ied particle physics models in the context of higher-dimensional gauge theories with non-commutative extra dimensions. These ideas have been developed in two complementary ways, namely (i) starting with a higher-dimensional gauge theory and dimensionally reducing it to four dimensions over fuzzy internal spaces and (ii) starting with a four-dimensional, renormalizable gauge theory and dynamically generating fuzzy extra dimensions. We describe the above approaches and moreover we discuss the inclusion of fermions and the construction of realistic chiral theories in this context.}


\Classification{70S15}

\tableofcontents

\renewcommand{\thefootnote}{\arabic{footnote}}
\setcounter{footnote}{0}

\section{Introduction}

The unif\/ication of the fundamental interactions has always been one of the main goals of theo\-retical physics. Several approaches have been employed in order to achieve this goal, one of the most exciting ones being the proposal that extra dimensions may exist in nature. The most serious support on the existence of extra dimensions came from superstring theories~\cite{Green:1987sp}, which
at present are the best candidates for a unif\/ied description of all
fundamental interactions, inclu\-ding gravity and moreover they can be
consistently def\/ined only in higher dimensions. Among superstring theories
the heterotic string \cite{Gross:1985fr} has always been considered as the most promising version in the prospect to f\/ind contact with low-energy physics studied in accelerators, mainly due to the presence of the ten-dimensional ${\cal N} = 1$ gauge sector. Upon compactif\/ication of the ten-dimensional space-time and subsequent dimensional reduction the initial $E_8 \times E_8$ gauge theory can break to phenomenologically interesting Grand Unif\/ied Theories (GUTs), where the Standard Model (SM) could in principle be accommodated~\cite{Gross:1985fr}. Dimensional
reduction of higher-dimensional gauge theories had been studied few years
earlier than the discovery of the heterotic superstring with pioneer
studies the Forgacs--Manton Coset Space Dimensional Reduction (CSDR)~\cite{Forgacs:1979zs,Kapetanakis:1992hf,Kubyshin:1989vd}
and the Scherk--Schwarz group manifold reduction~\cite{Scherk:1979zr}. In these frameworks gauge-Higgs unif\/ication is achieved in higher dimensions, since the four-dimensional gauge and
Higgs f\/ields are simply the surviving components of the gauge f\/ields of a pure gauge theory def\/ined in higher dimensions. Moreover in the CSDR the addition of fermions in the higher-dimensional gauge theory leads naturally to Yukawa couplings in four dimensions. A~major achievement in this direction is the possibility to obtain chiral theories in four dimensions~\cite{Manton:1981es}.

On the other hand, non-commutative geometry of\/fers another framework aiming to describe physics at the Planck scale \cite{Connes,Madore}. In
the spirit of non-commutative geometry also particle models with
non-commutative gauge theory were explored \cite{Connes:1990qp}
(see also~\cite{Martin:1996wh}),~\cite{DV.M.K., M.}. It is
worth stressing the observation that a natural realization of
non-commutativity of space appears in the string theory context of $D$-branes in the presence of a constant antisymmetric f\/ield \cite{Connes:1997cr}, which not only brought together the two approaches but they can be considered
complementary. Another interesting development in the non-commutative
framework was the work of Seiberg and Witten~\cite{SW}, where a map between the non-commutative and commutative gauge theories has been described. Based
on that and related subsequent developments~\cite{Chaic,Jurco:2000ja} a non-commutative
version of the SM has been constructed~\cite{SM}. These non-commutative models represent inte\-res\-ting
generalizations of the SM and hint at possible new physics.
However they do not address the usual problem of the SM, the
presence of a plethora of free parameters mostly related to the ad
hoc introduction of the Higgs and Yukawa sectors in the theory.

According to the above discussion it is natural to investigate higher-dimensional gauge theo\-ries and their dimensional reduction in four dimensions. Our aim is to provide an up to-date overview of certain attempts in this direction, developed over the last years. The development of these ideas has followed two complementary ways, namely (i)~the dimensional reduction of a~higher-dimensional gauge theory over fuzzy internal spaces~\cite{Aschieri:2003vy} and (ii)~the dynamical generation of fuzzy extra dimensions within a four-dimensional and renormalizable gauge theo\-ry~\cite{Aschieri:2006uw,Steinacker:2007ay,Chatzistavrakidis:2009ix,Chatzistavrakidis:2010xi}.

More specif\/ically, the paper is organized as follows. In Section~\ref{section2} we present a study of the CSDR in the
non-commutative context which sets the rules for constructing new
particle models that might be phenomenologically interesting. One
could study CSDR with the whole parent space $M^D$ being non-commutative or with just non-commutative Minkowski space or
non-commutative internal space. We specialize here to this last
situation and therefore eventually we obtain Lorentz covariant
theories on commutative Minkowski space. We further specialize to
fuzzy non-commutativity, i.e.\ to matrix type non-commutativity. Thus,
following~\cite{Aschieri:2003vy}, we consider non-commutative spaces like those
studied in~\cite{Madore,DV.M.K.,M.} and implementing the CSDR
principle on these spaces we obtain the rules for
constructing new particle models.  In Section~\ref{section2.1} the fuzzy sphere is introduced and moreover the gauge theory on the fuzzy sphere is discussed. In Section~\ref{section2.2} a trivial dimensional reduction of a higher-dimensional gauge theory over the fuzzy sphere is performed. In Section~\ref{section2.3} we discuss the non-trivial dimensional reduction; f\/irst the CSDR scheme in the commutative case is brief\/ly reviewed and subsequently it is applied to the case of fuzzy extra dimensions. In Section~\ref{section2.4} the issue of chirality is discussed within the above context.

In Section \ref{section3} we reverse the above approach \cite{Aschieri:2006uw} and
examine how a four-dimensional gauge theory dynamically develops
fuzzy extra dimensions.  In Sections~\ref{section3.1} and~\ref{section3.2} we present a~simple f\/ield-theoretical model which realizes the above ideas. It is
def\/ined as a renormalizable~$SU(N)$ gauge theory on four-dimensional
Minkowski space~$M^4$, containing three scalars in the adjoint of~$SU(N)$ that transform as vectors under an additional global~$SO(3)$
symmetry with the most general renormalizable potential.  We then
show that the model dynamically develops fuzzy extra dimensions,
more precisely a fuzzy sphere~$S^2_{N}$. The appropriate
interpretation is therefore as gauge theory on $M^4 \times S^2_{N}$.
The low-energy ef\/fective action is that of a four-dimensional gauge
theo\-ry on $M^4$, whose gauge group and f\/ield content is dynamically
determined by compactif\/ication and dimensional reduction on the
internal sphere  $S^2_{N}$.  An interesting and  rich pattern of
spontaneous symmetry breaking appears, namely the breaking of the original $SU(N)$
gauge symmetry down to
 either $SU(n)$ or $SU(n_1)\times
SU(n_2) \times U(1)$. The latter case is the generic one, and
implies also a monopole f\/lux induced on the fuzzy sphere. Moreover we determine explicitly the tower of massive Kaluza--Klein modes
corresponding to the ef\/fective geometry, which justif\/ies the
interpretation as a compactif\/ied higher-dimensional gauge theory.
Last but not least, the model is renormalizable. In Sections~\ref{section3.3} and~\ref{section3.4} we explore the dynamical generation of a pro\-duct
of two fuzzy spheres~\cite{Chatzistavrakidis:2009ix}.
Specif\/ically, we start with the $SU(N)$ Yang--Mills theory in four
dimensions, coupled to six scalars and four Majorana spinors,
i.e.\ with the particle spectrum of the $\cN=4$ supersymmetric
Yang--Mills theory (SYM). Adding an explicit $R$-symmetry-breaking
potential, thus breaking the $\cN=4$ supersymmetry,
we reveal stable $M^4\times S_L^2\times S^2_R$ vacua. In the most interesting case we include magnetic f\/luxes on the extra-dimensional fuzzy spheres and study the fermion spectrum, in particular the zero modes of the Dirac operator. The outcome of our analysis is that we obtain a mirror model in low energies.

In Section~\ref{section4} we present a recently developed approach within the above framework, which leads to chiral low-energy models \cite{Chatzistavrakidis:2010xi}. In particular, $\Z_3$ orbifolds of ${\cal N}=4$ supersymmetric Yang--Mills theory are discussed and they are subsequently used to dynamically generate fuzzy extra dimensions. The extra dimensions are described by twisted fuzzy spheres, def\/ined in Section~\ref{section4.2}. This framework allows to construct low-energy models with interesting unif\/ication groups and a chiral spectrum. In particular, we are led to study three dif\/ferent models based on the gauge groups $SU(4)\times SU(2)\times SU(2)$, $SU(4)^3$ and $SU(3)^3$ respectively. The spontaneous symmetry breaking of the latter unif\/ied gauge group down to the minimal supersymmetric standard model and to the $SU(3)\times U(1)_{em}$ is subsequently studied within the same framework. Finally, Section~\ref{section5} contains our conclusions.

\section{Fuzzy spaces and dimensional reduction}\label{section2}

\subsection{The fuzzy sphere}\label{section2.1}

The fuzzy sphere \cite{Madore:1991bw} is a noncommutative manifold which corresponds to a matrix approximation of the ordinary sphere. In order to describe it let us consider the ordinary sphere as a~submanifold of the three-dimensional Euclidean space $\R^3$. The coordinates of $\R^3$ will be denoted as $x_a$, $a=1,2,3$. Then the algebra of functions on the ordinary sphere $S^2\subset \R^3$ can be generated by the coordinates of $\R^3$ modulo the relation
 \[
 \sum_{a=1}^3 x_ax_a=R^2,
  \]
  where $R$ is the radius of the sphere. Clearly, the sphere admits the action of a global $SO(3) \sim SU(2)$ isometry group. The generators of $SU(2) \sim SO(3)$ are the three angular momentum
operators $L_a$,
\[
L_{{a}} = -i \varepsilon_{{a}{b}{c}} x_{{b}}
\partial_{{c}},
\]
which in terms of the usual spherical coordinates $\theta$ and $\phi$ become
\begin{gather*}
L_1  =  i \sin \phi
\frac{\partial}{\partial \theta} + i \cos \phi
\cot \theta
\frac{\partial}{\partial \phi}, \\
L_2  =  -i  \cos \phi  \frac{\partial}{\partial
\theta} + i \sin \phi \cot \theta
\frac{\partial}{\partial \phi}, \\
L_3  =  -i  \frac{\partial}{\partial \phi}.
\end{gather*}
These relations can be summarized as
\begin{gather*}
L_{{a}} = -i \xi^{\alpha}_{{a}} \partial_{\alpha},
\end{gather*} where the Greek index $\alpha$ corresponds to the spherical coordinates and $\xi_a^{\alpha}$ are the components of the Killing vector f\/ields associated with the isometries of the sphere.
The metric tensor of the sphere can be expressed in terms of the Killing
vectors as
\[
g^{\alpha\beta} = \frac{1}{R^2}   \xi^{\alpha}_{{a}} \xi^{\beta}_{{a}}.
\]

Any function on the sphere can be expanded in terms of the
eigenfunctions of the sphere,
\begin{gather}\label{eq:expand}
a(\theta, \phi) = \sum^{\infty}_{l=0} \sum^l_{m=-l} a_{lm}
Y_{lm}(\theta, \phi),
\end{gather}
where $a_{lm}$ is a complex coef\/f\/icient and $Y_{lm}(\theta, \phi)$
are the spherical harmonics, which satisfy the equation
\[
L^2 Y_{lm} = -R^2 \Delta_{S^2} Y_{lm} = l(l+1) Y_{lm},
\]
where $\Delta_{S^2}$ is the scalar Laplacian on the sphere
\[
\Delta_{S^2} = \frac{1}{\sqrt{g}}   \partial_a \big( g^{ab} \sqrt{g}
\partial_b\big).
\]
The spherical harmonics have an eigenvalue $\mu \sim l(l+1)$ for
integer \mbox{$l = 0,1, \dots$}, with degeneracy $2l+1$. The
orthogonality condition of the spherical harmonics is
\[
\int d\Omega   Y^{\dag}_{lm} Y^{\phantom{\dag}}_{l'm'} = \delta_{l
l'}   \delta_{m m'},
\]
where $d\Omega = \sin \theta   d\theta d\phi$.

The spherical harmonics can be expressed in terms of the cartesian
coordinates $x_{{a}}$ of a unit vector in
$\R^{3}$,
\begin{gather}\label{eq:spharm}
Y_{lm}(\theta, \phi) = \sum_{\vec{a}} f ^{(lm)} _{{a_{1}} \dots
{a_{l}}} x^{{a_{1}}}  \cdots x^{{a_{l}}}
\end{gather}
where $f ^{(lm)} _{{a_{1}} \dots {a_{l}}}$ is a traceless
symmetric tensor of $SO(3)$ with rank~$l$.

Similarly we can expand $N \times N$ matrices on a sphere as,
\begin{gather}\label{eq:fuzzyexpand}
\hat{a}  =  \sum^{N-1}_{l=0} \sum^l_{m=-l} a_{lm} \hat{Y}_{lm}, \qquad
\hat{Y}_{lm}
 =  R^{-l}\sum_{\vec{a}} f ^{(lm)} _{{a_{1}} \dots
{a_{l}}} \hat{X}^{{a_{1}}}  \cdots \hat{X}^{{a_{l}}},
\end{gather}
where
\begin{gather}
\label{coordfnct} \hat{X}_{{a}}=\frac{2R}{\sqrt{N^{2}-1}}
\lambda^{(N)}_{{a}}
\end{gather}
 and $\lambda^{(N)}_a$ are the generators of $SU(2)$ in the
$N$-dimensional representation. The tensor $f ^{(lm)}_{\hat{a_{1}} \dots
\hat{a_{l}}}$ is the same one as in~(\ref{eq:spharm}). The
matrices $\hat{Y}_{lm}$ are known as fuzzy spherical harmonics for
reasons which will be apparent shortly. They obey the
orthonormality condition
\[
\textrm{Tr}_N \left( \hat{Y}^{\dag}_{lm}
\hat{Y}_{l'm'}^{\phantom{\dag}} \right) = \delta_{l l'}   \delta_{m
m'}.
\]
There is an obvious relation between equations (\ref{eq:expand}) and
(\ref{eq:fuzzyexpand}), namely\footnote{Let us note that in general the map from matrices to functions is not unique, since the expansion coef\/f\/i\-cients~$a_{lm}$ may be dif\/ferent. However, here we introduce the fuzzy sphere by truncating the algebra of functions on the ordinary sphere and therefore the use of the same expansion coef\/f\/icients is a natural choice.}
\begin{gather*}
\hat{a} = \sum^{N-1}_{l=0} \sum^{l}_{m=-l} a_{lm} \hat{Y}_{lm} \quad \to  \quad
a(\theta, \phi) = \sum^{N-1}_{l = 0} \sum^l_{m = -l} a_{lm}
Y_{lm}(\theta, \phi).
\end{gather*}
Notice that the expansion in spherical harmonics is truncated at
$N-1$ ref\/lecting the f\/inite number of degrees of freedom in the
matrix~$\hat{a}$. This allows the consistent def\/inition of a matrix
approximation of the sphere known as fuzzy sphere.

According to
the above discussion the fuzzy sphere~\cite{Madore:1991bw} is a matrix
approximation of the usual sphere~$S^2$. The algebra of functions on~$S^2$ (for example spanned by the spherical harmonics) is truncated at a given frequency and thus
becomes f\/inite-dimensional. The truncation has to be consistent with
the associativity of the algebra and this can be nicely achieved
relaxing the commutativity property of the algebra. The fuzzy sphere
is the ``space'' described by this non-commutative algebra. The
algebra itself is that of $N \times N$ matrices, which we denote as ${\rm Mat}(N;\C)$. More precisely, the fuzzy
sphere $S^2_{N}$ at fuzziness level $N-1$ is the non-commutative
manifold whose coordinate functions $\hat{X}_{{a}}$ are $N \times
N$ hermitian matrices proportional to the generators of the
$N$-dimensional representation of $SU(2)$ as in equation~(\ref{coordfnct}). They satisfy the
condition $ \sum_{{a}=1}^{3} \hat X_{{a}} \hat X_{{a}} =
R^{2}$ and the commutation relations
\[
[ \hat X_{{a}}, \hat X_{{b}} ] =i\alpha \varepsilon_{{a} {b} {c}}
\hat X_{{c}},
\]
where $\alpha=\frac{2R}{\sqrt{N^{2}-1}}$. It can be proven that for $N\to
\infty$ one obtains the usual commutative sphere. In the following we shall mainly work with the following antihermitian matrices,
 \[
 X_a=\frac{\hat X_a}{i\alpha R},
  \]
  which describe equivalently the algebra of the fuzzy sphere and they satisfy the relations
\begin{gather*}
\sum_{a=1}^3 X_a X_a  =  -\frac{1}{\alpha^2}, \qquad
 [ X_{a}, X_{b} ]  =  C_{abc}X_c,
\end{gather*}
  where $C_{{a} {b} {c}}= \varepsilon_{{a}{b}{c}}/R$.

On the fuzzy sphere there is a natural $SU(2)$ covariant
dif\/ferential calculus. This calculus is three-dimensional and the
derivations $e_{{a}}$ along $X_{{a}}$ of a function $ f$ are
given by $e_{{a}}({f})=[X_{{a}}, {f}]$. 
Accordingly the action of the Lie derivatives on functions is given
by
\begin{gather*}
{\cal L}_{{a}} f = [{X}_{{a}},f ];
\end{gather*}
these Lie derivatives satisfy the Leibniz rule and the $SU(2)$ Lie
algebra relation
\begin{gather*}
[ {\cal L}_{{a}}, {\cal L}_{{b}} ] = C_{{a} {b}
{c}} {\cal L}_{{c}}.
\end{gather*}
In the $N \to \infty$ limit the derivations $e_{{a}}$
become $
e_{{a}} = C_{{a}{b}{c}} x^{{b}}
\partial^{{c}}
$ and only in this commutative limit the tangent space becomes
two-dimensional. The exterior derivative is given by
\[
d f = [X_{{a}},f]\theta^{{a}}
\]
with $\theta^{{a}}$ the one-forms dual to the vector f\/ields
$e_{{a}}$,
$\langle e_{{a}},\theta^{{b}}\rangle =\delta_{{a}}^{{b}}$. The
space of one-forms is generated by the $\theta^{{a}}$'s in the
sense that for any one-form $\omega=\sum_i f_i d h_i \:t_i$ we can
always write
$\omega=\sum_{{a}=1}^3{\omega}_{{a}}\theta^{{a}}$ with
given functions $\omega_{{a}}$ depending on the functions $f_i$,
$h_i$ and $t_i$. The action of the Lie derivatives ${\cal
L}_{{a}}$ on the one-forms $\theta^{{b}}$ explicitly reads
\begin{gather*}
{\cal L}_{{a}}(\theta^{{b}}) =  C_{{a}{b}{c}}
\theta^{{c}}.
\end{gather*}
On a general one-form $\omega=\omega_{{a}}\theta^{{a}}$ we
have $ {\cal L}_{{b}}\omega={\cal
L}_{{b}}(\omega_{{a}}\theta^{{a}})=
\left[X_{{b}},\omega_{{a}}\right]\theta^{{a}}-\omega_{{a}}C^{{a}}_{\
{b} {c}}\theta^{{c}} $ and therefore
\begin{gather*}
({\cal
L}_{{b}}\omega)_{{a}}=\left[X_{{b}},\omega_{{a}}\right]-
\omega_{{c}}C^{{c}}_{\ {b}  {a}}.
\end{gather*}

The dif\/ferential geometry on  the product space Minkowski times
fuzzy sphere, $M^{4} \times S^2_{N}$, is easily obtained from that
on $M^4$ and on $S^2_N$. For example a one-form $A$ def\/ined on
$M^{4} \times S^2_{N}$ is written as
\begin{gather*}
A= A_{\mu} dx^{\mu} + A_{{a}} \theta^{{a}}
\end{gather*}
with $A_{\mu} =A_{\mu}(x^{\mu}, X_{{a}} )$ and $A_{{a}}
=A_{{a}}(x^{\mu}, X_{{a}} )$.

One can also introduce spinors on the fuzzy sphere and study the Lie
derivative on these spinors~\cite{Aschieri:2003vy}. Although here we have sketched the
dif\/ferential geometry on the fuzzy sphere, one can study other
(higher-dimensional) fuzzy spaces (e.g.\ fuzzy~$CP^M$~\cite{Balachandran:2001dd}, see also~\cite{Trivedi:2000mq}) and with
similar techniques their dif\/ferential geometry.

\subsubsection{Gauge theory on the fuzzy sphere}

In order to describe gauge f\/ields on the fuzzy sphere it is natural to introduce the notion of covariant coordinates~\cite{Madore:2000en}. In order to do so let us begin with a f\/ield $\phi(X_a)$ on the fuzzy sphere, which is a polynomial in the $X_a$ coordinates.  An inf\/initesimal gauge transformation
$\delta\phi$ of the f\/ield $\phi$ with gauge transformation
parameter $\lambda(X_a)$ is def\/ined by
\[
\delta\phi(X) = \lambda(X)\phi(X).
\]
This is an inf\/initesimal Abelian $U(1)$ gauge transformation if
$\lambda(X)$ is just an antihermitian function of the coordinates
$X_a$, while it is an inf\/initesimal non-Abelian $U(P)$ gauge
transformation if $\lambda(X)$ is valued in ${\rm{Lie}}(U(P))$,
the Lie algebra of hermitian $P\times P$ matrices. In the
following we will always assume ${\rm{Lie}}(U(P))$ elements to
commute with the coordina\-tes~$X_a$. The coordina\-tes~$X_a$ are
invariant under a gauge transformation
\[
\delta X_{{a}} = 0.
\]
Then, multiplication of a f\/ield on the left by a coordinate is not
a covariant operation in the non-commutative case. That is
\[
\delta(X_{{a}}\phi) = X_{{a}}\lambda(X)\phi,
\]
and in general the right hand side is not equal to
$\lambda(X)X_{a}\phi$. Following the ideas of ordinary gauge
theory one then introduces covariant coordinates $\phi_{{a}}$
such that
\begin{gather*}
\delta(\phi_{{a}}{}\phi) = \lambda\phi_{{a}}{}\phi.
\end{gather*}
This happens if
\begin{gather}
\label{covtr}
\delta(\phi_{{a}})=[\lambda,\phi_{{a}}].
\end{gather}
The analogy with ordinary gauge theory also suggests to set
\[
\phi_{{a}} \equiv X_{{a}} + A_{{a}}
\]
and interpret $A_a$ as the gauge potential of the non-commutative
theory. Then $\phi_a$ is the non-commutative analogue of a
covariant derivative. The transformation properties of
$A_{{a}}$ support the interpretation of $A_a$ as gauge f\/ield, since from requirement (\ref{covtr}) we can deduce that $A_a$ transforms as
\[
\delta A_{{a}} = -[ X_{{a}}, \lambda] +
[\lambda,A_{{a}}].
\]
Correspondingly we can def\/ine a tensor $F_{{a}{b}}$,
the analogue of the f\/ield strength, as
\begin{gather}
\label{2.33}
F_{{a} {b}} = [ X_{{a}}, A_{{b}}] -
[ X_{{b}}, A_{{a}} ] + [A_{{a}} , A_{{b}} ] -
C^{{c}}_{\ {a} {b}}A_{{c}}  =  [ \phi_{{a}},
\phi_{{b}}] - C^{{c}}_{\ {a} {b}}\phi_{{c}}.
\end{gather}
The presence of the last term in (\ref{2.33}) might seem strange at f\/irst sight, however it is imposed in the def\/inition of the f\/ield strength by the requirement of covariance. Indeed, it is straightforward to prove that the above tensor transforms covariantly, i.e.
\[
\delta F_{{a} {b}} = [\lambda, F_{{a} {b}}].
\]
Similarly, for a spinor $\psi$ in the adjoint representation, the
inf\/initesimal gauge transformation is given by
\[
\delta \psi = [\lambda, \psi].
\]

\subsection{Dimensional reduction and gauge symmetry enhancement}\label{section2.2}

Let us now consider a non-commutative gauge theory on
$M^{4} \times (S/R)_{F}$ with gauge
group $G=U(P)$ and examine its four-dimensional interpretation.
$(S/R)_{F}$ is a fuzzy coset, for example the fuzzy sphere
$S^{2}_{N}$. The action is
\begin{gather}\label{formula8}
{\cal S}_{\rm YM}=\frac{1}{4g^{2}} \int d^{4}x\, k{\rm Tr}\, {\rm tr}_{G}\,
F_{MN}F^{MN},
\end{gather}
where $k{\rm Tr}$ denotes integration over the fuzzy coset $(S/R)_F$
described by $N\times N$ matrices; here the parameter $k$ is related
to the size of the fuzzy coset space. For example for the fuzzy
sphere we have $R^{2} = \sqrt{N^{2}-1}\pi k$~\cite{Madore}. In the
$N\to \infty$ limit $k{\rm Tr}$ becomes the usual integral on the
coset space. For f\/inite $N$, ${\rm Tr}$ is a good integral because it has
the cyclic property ${\rm Tr}(f_1\cdots f_{p-1}f_p)={\rm Tr}(f_pf_1\cdots
f_{p-1})$. It is also invariant under the action of the group $S$,
that is  inf\/initesimally given by the Lie derivative. In the action~(\ref{formula8}) ${\rm tr}_G$ is the gauge group $G$ trace. The
higher-dimensional f\/ield strength $F_{MN}$, decomposed in
four-dimensional space-time and extra-dimensional components, reads
as $(F_{\mu \nu}, F_{\mu {b}}, F_{{a} {b} })$, where $\mu$, $\nu$ are four-dimensional spacetime indices. The various components of the f\/ield strength are explicitly given by
\begin{gather*}
F_{\mu \nu}  =
\partial_{\mu}A_{\nu} -
\partial_{\nu}A_{\mu} + [A_{\mu}, A_{\nu}],\\
F_{\mu {a}}  =
\partial_{\mu}A_{{a}} - [X_{{a}}, A_{\mu}] + [A_{\mu},
A_{{a}}],  \\
F_{{a} {b}} =   [ X_{{a}}, A_{{b}}] - [
X_{{b}}, A_{{a}} ] + [A_{{a}} , A_{{b}} ] -
C^{{c}}_{\ {a} {b}}A_{{c}}.
\end{gather*} In terms
of the covariant coordinates $\phi$, which were introduced in the previous section, the f\/ield strength in the non-commutative directions
becomes
\begin{gather*}
F_{\mu {a}} =
\partial_{\mu}\phi_{{a}} + [A_{\mu}, \phi_{{a}}]=
D_{\mu}\phi_{{a}},\\
F_{{a} {b}}  =  [\phi_{{a}}, \phi_{{b}}] -
C^{{c}}_{\ {a} {b}} \phi_{{c}}.
\end{gather*}
Using these expressions the action (\ref{formula8}) becomes
\begin{gather}
{\cal S}_{\rm YM}= \int d^{4}x\, {\rm Tr}\, {\rm tr}_{G}\,\left( \frac{k}
{4g^{2}}F_{\mu \nu}^{2} + \frac{k}{2g^{2}}(D_{\mu}\phi_{{a}})^{2} -
V(\phi)\right),\label{theYMaction}
\end{gather}
where the potential term $V(\phi)$ is the $F_{{a} {b}}$
kinetic term (in our conventions $F_{{a} {b}}$ is
antihermitian so that $V(\phi)$ is hermitian and non-negative),
\begin{gather}
V(\phi)=-\frac{k}{4g^{2}}\,{\rm  Tr}\,{\rm tr}_G \sum_{{a} {b}}
F_{{a} {b}} F_{{a} {b}}
\nonumber \\
\phantom{V(\phi)=}{}
=-\frac{k}{4g^{2}} \,{\rm Tr}\,{\rm tr}_G \left( [\phi_{{a}},
\phi_{{b}}][\phi^{{a}}, \phi^{{b}}] -
4C_{{a} {b} {c}} \phi^{{a}} \phi^{{b}}
\phi^{{c}} + 2R^{-2}\phi^{2} \right).\label{pot1}
\end{gather}
The action (\ref{theYMaction}) is naturally interpreted as an action
in four dimensions. The inf\/initesimal $G$ gauge transformation with
gauge parameter $\lambda(x^{\mu},X^{{a}})$ can indeed be
interpreted just as an $M^4$ gauge transformation. We write
\begin{gather}
\lambda(x^{\mu},X^{{a}})=\lambda^{I}(x^{\mu},X^{{a}}){\cal
T}^{I} =\lambda^{h, I}(x^{\mu})T^{h}{\cal
T}^{I},\label{3.33}
\end{gather}
where ${\cal T}^{I}$ are hermitian generators of $U(P)$,
$\lambda^{I}(x^\mu,X^{{a}})$ are $N\times N$ antihermitian
matrices and thus are expressible as $\lambda(x^\mu)^{I ,
h}T^{h}$, where $T^{h}$ are antihermitian generators of $U(N)$. The
f\/ields~$\lambda(x^{\mu})^{I , h}$, with $h=1,\ldots, N^2$, are
the Kaluza--Klein modes of $\lambda(x^{\mu}, X^{{a}})^{I}$.
We now consider on equal footing the indices $h$ and $I$ and
interpret the f\/ields on the r.h.s.\ of (\ref{3.33}) as one f\/ield
valued in the tensor product Lie algebra ${\rm{Lie}}(U(N)) \otimes
{\rm{Lie}}(U(P))$. This Lie algebra is indeed ${\rm{Lie}}(U(NP))$
(the $(NP)^2$ generators $T^{h}{\cal T}^{I}$ being $NP\times
NP$ antihermitian matrices that are linear independent). Similarly
we rewrite the gauge f\/ield $A_\nu$ as
\[
A_\nu(x^{\mu},X^{{a}})=A_{\nu}^{I}(x^{\mu},X^{{a}}){\cal
T}^{I} =A_{\nu}^{h, I}(x^{\mu})T^{h}{\cal T}^{I},
\]
and interpret it as a ${\rm{Lie}}(U(NP))$-valued gauge f\/ield on
$M^4$. The four-dimensional scalar f\/ields $\phi_{{a}}$ are interpreted similarly. It is worth noting that the scalars transform in the adjoint representation of the four-dimensional gauge group and therefore they are not suitable for the electroweak symmetry breaking. This serves as a motivation to use a non-trivial dimensional reduction scheme, which is presented in the following section. Finally ${\rm Tr}\, {\rm tr}_{G}$
is the trace over~$U(NP)$ matrices in the fundamental
representation.

Up to now we have just performed a ordinary fuzzy dimensional
reduction. Indeed in the commutative case the expression
(\ref{theYMaction}) corresponds to rewriting the initial lagrangian
on $M^4\times S^2$ using spherical harmonics on $S^2$. Here the
space of functions is f\/inite-dimensional and therefore the inf\/inite
tower of modes reduces to the f\/inite sum given by the trace $\rm Tr$. The remarkable result of the above analysis is that the gauge group in four dimensions is enhanced compared to the gauge group $G$ in the higher-dimensional theory. Therefore it is very interesting to note that we can in fact start with an Abelian gauge group in higher dimensions and obtain non-Abelian gauge symmetry in the four-dimensional theory.

\subsection{Non-trivial dimensional reduction over fuzzy
extra dimensions}\label{section2.3}

In this section we reduce the number of gauge f\/ields and
scalars in the action (\ref{theYMaction}) by applying the Coset
Space Dimensional Reduction (CSDR) scheme. Before proceeding to the case of fuzzy extra dimensions let us brief\/ly recall how this scheme works in the commutative case.

\subsubsection{Ordinary CSDR}

One way to dimensionally reduce a gauge theory on $M^{4} \times
S/R$ with gauge group $G$ to a gauge theory on $M^4$, is to
consider f\/ield conf\/igurations that are invariant under $S/R$
transformations. Since the action of the group $S$ on the coset
space $S/R$ is transitive (i.e., connects all points), we can
equivalently require the f\/ields in the theory to be invariant
under the action of $S$ on $S/R$. Inf\/initesimally, if we denote by
$\xi_{a}$ the Killing vectors on $S/R$ associated to the
generators~$T^a$ of~$S$, we require the f\/ields to have zero Lie
derivative along $\xi_{a}$. For scalar f\/ields this is
equivalent to requiring independence under the $S/R$ coordinates.
The CSDR scheme dimensionally reduces a gauge theory on $M^{4}
\times S/R$ with gauge group $G$ to a gauge theory on $M^4$
imposing a milder constraint, namely the f\/ields are required to be
invariant under the $S$ action up to a $G$ gauge transformation
\cite{Forgacs:1979zs,Kapetanakis:1992hf,Kubyshin:1989vd}\footnote{See also \cite{Lechtenfeld:2006wu} for related work.}. Thus we have,
respectively for scalar f\/ields $\phi$ and the one-form gauge f\/ield
$A$,
\begin{gather}
 {\cal L}_{\xi_a}\phi=\delta^{W_a}\phi=W_a\phi,\qquad {\cal L
}_{\xi_a}A=\delta^{W_a}A=-DW_a ,\label{condCSDR}
\end{gather}
where $\delta^{W_a}$ is the inf\/initesimal gauge transformation
relative to the gauge parameter ${W_a}$ that depends on the coset
coordinates (in our notations  $A$ and $W_a$  are
antihermitian and the covariant derivative reads $D=d+A$).
The gauge parameters ${W_a}$ obey a consistency condition which
follows from the relation
\begin{gather}\label{cc}
[{\cal L}_{\xi_{{a}}},{\cal L}_{\xi_{{b}}}] = {\cal
L}_{[\xi_{{a}},\xi_{{b}}]}
\end{gather}
and transform under a gauge transformation $\phi\to g\phi$
as
\begin{gather}\label{W}
W_{{a}} \to gW_{{a}}g^{-1} +
({\cal L}_{\xi_{{a}}}g)g^{-1}.
\end{gather}
Since two points of the coset are connected by an
$S$-transformation which is equivalent to a gauge transformation,
and since the Lagrangian is gauge invariant, we can study the
above equations just at one point of the coset, let's say $y^\a=0$,
where we denote by $(x^\mu,y^\a)$ the coordinates of $M^4\times
S/R$, and we use $ a$, $\a$, $i$ to denote $S$, $S/R$ and $R$ indices. In
general, using (\ref{W}),  not all the $W_{{a}}$ can be gauged
transformed to zero at $y^\a=0$, however one can choose $W_{\a} =0$
denoting by $W_{i}$ the remaining ones. Then the consistency
condition which follows from equation~(\ref{cc}) implies that~$W_{i}$
are constant and equal to the generators of the embedding of~$R$
in~$G$ (thus in particular~$R$ must be embeddable in $G$; we write~$R_G$ for the image of $R$ in $G$).

The detailed analysis of the constraints given in~\cite{Forgacs:1979zs,Kapetanakis:1992hf} provides us with the
four-dimensional unconstrained f\/ields as well as with the gauge
invariance that remains in the theory after dimensional reduction.
Here we just state the results:
\begin{itemize}
\itemsep=0pt
 \item The components $A_{\mu}(x,y)$ of the
initial gauge f\/ield $A_{M}(x,y)$ become, after dimensional
reduction, the four-dimensional gauge f\/ields and furthermore they
are independent of $y$. In addition one can f\/ind that they have to
commute with the elements of the $R_{G}$ subgroup of~$G$. Thus the
four-dimensional gauge group~$H$ is the centralizer of~$R$ in $G$,
$H=C_{G}(R_{G})$.
\item Similarly, the $A_\a(x,y)$ components of
$A_{M}(x,y)$ denoted by $\phi_\a(x,y)$ from now on, become scalars
in four dimensions. These f\/ields transform under $R$ as a vector
$v$, i.e.
\begin{gather*}
S \supset R,\\ {\rm adj}\, S  = {\rm  adj}\, R+v .
\end{gather*}
Moreover $\phi_\a(x,y)$ acts as an intertwining operator connecting
induced representations of $R$ acting on $G$ and $S/R$. This
implies, exploiting Schur's lemma, that the transformation
properties of the f\/ields $\phi_\a(x,y)$ under $H$ can be found if
we express the adjoint representation of $G$ in terms of $R_{G}
\times H$:
\begin{gather*}
G  \supset  R_{G} \times H, \nonumber \\
{\rm adj}\, G  = ({\rm adj}\, R,1)+(1,{\rm adj}\, H)+\sum(r_{i},h_{i}).
\end{gather*}
Then if $v=\sum s_{i}$, where each $s_{i}$ is an irreducible
representation of $R$, there survives an~$h_{i}$ multiplet for
every pair $(r_{i},s_{i})$, where $r_{i}$ and $s_{i}$ are
identical irreps.\ of~$R$. If we start from a~pure gauge theory on
$M^4\times S/R$, the four-dimensional potential (at $y^\a=0$) can
be shown to be given by
\begin{gather*}
V=- \frac{1}{4} F_{\a\b}F^{\a\b}= -\frac{1}{4} (C^{{c}}_{\
\a\b}\phi_{{c}} - [\phi_{\a}, \phi_{\b}])^{2},
\end{gather*}
where we have def\/ined  $\phi_{i}\equiv W_{i}$. However, the f\/ields
$\phi_\a$ are not independent because the conditions~(\ref{condCSDR}) at $y^\a=0$ constrain them. The solution of the
constraints provides the physical dimensionally reduced f\/ields in
four dimensions; in terms of these physical f\/ields the potential
is still a quartic polynomial. Then, the minimum of this potential
will determine the spontaneous symmetry breaking pattern.
\item Turning next to the fermion f\/ields, similarly to scalars, they act
as an intertwining operator connecting induced representations of
$R$ in $G$ and in $SO(d)$, where~$d$ is the dimension of the
tangent space of~$S/R$. Proceeding along similar lines as in the
case of scalars, and considering the more interesting case of even
dimensions, we impose f\/irst the Weyl condition. Then to obtain the
representation of $H$ under which the four-dimensional fermions
transform, we have to decompose the fermion representation
$F$ of the initial gauge group $G$ under $R_{G} \times H$,
i.e.
\begin{gather*}
F= \sum (t_{i},h_{i}),
\end{gather*}
and the spinor of $SO(d)$ under $R$
\begin{gather*}
\sigma_{d} = \sum \sigma_{j}.
\end{gather*}
Then for each pair $t_{i}$ and $\sigma_{i}$, where $t_{i}$ and
$\sigma_{i}$ are identical irreps.\ there is an $h_{i}$ multiplet
of spinor f\/ields in the four-dimensional theory. In order however
to obtain chiral fermions in the ef\/fective theory we have to
impose further requirements~\cite{Kapetanakis:1992hf,Manton:1981es}. The issue of chiral fermions will be discussed in Section~\ref{section2.4}.
\end{itemize}

\subsubsection{Fuzzy CSDR}

Let us now discuss how the above scheme can be applied in the case where the extra dimensions are fuzzy coset spaces \cite{Aschieri:2003vy}\footnote{A similar approach has also been considered in \cite{Harland:2009kg}.}. Since $SU(2)$ acts on the
fuzzy sphere $(SU(2)/U(1))_F$, and more in general the group $S$
acts on the fuzzy coset $(S/R)_F$, we can state the CSDR principle
in the same way as in the continuum case, i.e.\ the f\/ields in the
theory must be invariant under the inf\/initesimal $SU(2)$,
respectively $S$, action up to an inf\/initesimal gauge transformation
\begin{gather}\label{c1}
{\cal L}_{{b}} \phi = \delta_{W_{{b}}}\phi= W_{{b}}\phi,
\\
{\cal L}_{{b}}A = \delta_{W_{{b}}}A=-DW_{{b}},
\label{csdr}
\end{gather}
where $A$ is the one-form gauge potential $A = A_{\mu}dx^{\mu} +
A_{{a}} \theta^{{a}}$, and $W_{{b}}$ depends only on the
coset coordinates $X^{{a}}$ and (like $A_\mu$, $A_a$) is
antihermitian. We thus write $W_{{b}}=W_{{b}}^{I}{\cal
T}^{I}$, $I=1,2,\ldots, P^2,$ where ${\cal  T}^I$ are
hermitian generators of $U(P)$ and $(W_b^I)^\dagger=-W_b^I$; here
${}^\dagger$ is hermitian conjugation on the $X^{{a}}$'s.

In terms of the covariant coordinate $\phi_{{a}} =X_{{a}}
+ A_{{a}}$ and of
\[
\omega_{{a}} \equiv X_{{a}} - W_{{a}},
\]
the CSDR constraints (\ref{c1}) and (\ref{csdr}) assume a particularly simple form, namely
\begin{gather}\label{3.19}
[\omega_{{b}}, A_{\mu}] =0,
\\
\label{eq7}
C_{{b} {d} {e}} \phi^{{e}} = [\omega_{{b}},
\phi_{{d}} ].
\end{gather}
In addition we  have a consistency condition  following from the
relation $[{\cal{L}}_{{a}},{\cal{L}}_{{b}}]=
C_{{a}{b}}^{~~{c}}{\cal{L}}_{{c}}$:
\begin{gather}\label{3.17}
[ \omega_{{a}} , \omega_{{b}}] = C_{{a} {b}}^{\ \
{c}} \omega_{c},
\end{gather}
where $\omega_{{a}}$ transforms as $ \omega_{{a}}\to
\omega'_{{a}} = g\omega_{{a}}g^{-1}. $ One proceeds in a
similar way for the spinor f\/ields~\cite{Aschieri:2003vy}.

\subsubsection{Solving the CSDR constraints for
the fuzzy sphere}\label{section2.3.3}

We consider $(S/R)_{F}=S^{2}_{N}$, i.e.\ the fuzzy
sphere, and to be def\/inite at fuzziness level $N-1$ ($N \times N$
matrices). We study here the basic example where the gauge group is
$G=U(1)$. In this case the
$\omega_{{a}}=\omega_{{a}}(X^{{b}})$ appearing in the
consistency condition (\ref{3.17}) are $N \times N$ antihermitian
matrices and therefore can be interpreted as elements of
${\rm{Lie}}(U(N))$. On the other hand the $\omega_{{a}}$ satisfy
the commutation relations (\ref{3.17}) of ${\rm{Lie}}(SU(2))$.
Therefore in order to satisfy the consistency condition (\ref{3.17})
we have to embed ${\rm{Lie}}(SU(2))$ in ${\rm{Lie}}(U(N))$. Let
$T^h$ with $h = 1, \ldots ,(N)^{2}$ be the generators of
${\rm{Lie}}(U(N))$ in the fundamental representation. Then we can always
use the convention $h= ({a} , u)$ with ${a} = 1,2,3$ and $u=
4,5,\ldots, N^{2}$ where the $T^{{a}}$ satisfy the $SU(2)$ Lie
algebra,
\begin{gather}\label{su2com}
[T^{{a}}, T^{{b}}] = C^{{a} {b}}_{\ \
{c}}T^{{c}}.
\end{gather}
Then we def\/ine an embedding by identifying
\begin{gather}
 \omega_{{a}}= T_{{a}}.
\label{embedding}
\end{gather}
The constraint (\ref{3.19}), $[\omega_{{b}} , A_{\mu}] = 0$,
then implies that the four-dimensional gauge group $K$ is the
centralizer of the image of $SU(2)$ in $U(N)$, i.e. \[
K=C_{U(N)}(SU((2))) = SU(N-2) \times U(1)\times U(1),\]   where the
last $U(1)$ is the $U(1)$ of $U(N)\simeq SU(N)\times U(1)$. The
functions $A_{\mu}(x,X)$ are arbitrary functions of $x$ but the $X$
dependence is such that $A_{\mu}(x,X)$ is ${\rm{Lie}}(K)$-valued
instead of ${\rm{Lie}}(U(N))$, i.e. eventually we have a
four-dimensional gauge potential $A_\mu(x)$ with values in
${\rm{Lie}}(K)$. Concerning the constraint (\ref{eq7}), it is
satisf\/ied by choosing
\begin{gather}
\label{soleasy} \phi_{{a}}=r \phi(x) \omega_{{a}},
\end{gather}
i.e. the unconstrained degrees of freedom correspond to the scalar
f\/ield $\phi(x)$ which is a singlet under the four-dimensional
gauge group $K$.

The choice (\ref{embedding}) def\/ines one of the possible embedding
of ${\rm{Lie}}(SU(2))$ in ${\rm{Lie}}(U(N))$. For example, we could
also embed ${\rm{Lie}}(SU(2))$ in ${\rm{Lie}}(U(N))$ using the
irreducible $N$-dimensional rep.\ of $SU(2)$, i.e.\ we could identify
$\omega_{{a}}= X_{{a}}$. The constraint (\ref{3.19}) in this
case implies that the four-dimensional gauge group is $U(1)$ so that
$A_\mu(x)$ is $U(1)$ valued. The constraint~(\ref{eq7}) leads again
to the scalar singlet $\phi(x)$.

In general, we start with a $U(1)$ gauge theory on $M^4\times
S^2_N$. We solve the CSDR constraint~(\ref{3.17}) by embedding
$SU(2)$ in $U(N)$. There exist $p_{N}$ embeddings, where $p_N$ is
the number of ways one can partition the integer $N$ into a set of
non-increasing positive integers~\cite{Madore:1991bw}. Then the constraint
(\ref{3.19}) gives the surviving four-dimensional gauge group. The
constraint~(\ref{eq7}) gives the surviving four-dimensional scalars
and equation~(\ref{soleasy}) is always a~solution but in general not the
only one. By setting $\phi_{{a}}=\omega_{{a}}$ we obtain
always a minimum of the potential. This minimum is given by the
chosen embedding of $SU(2)$ in $U(N)$.

Concerning fermions in the adjoint, the corresponding analysis in~\cite{Aschieri:2003vy} shows that we have to consider the embedding \[ S \subset SO(\dim S), \]
which is given by $T_{{a}} = \frac{1}{2}C_{{a} {b}
{c}} \Gamma^{{b} {c}}$ that satisf\/ies the commutation relation \eqref{su2com}. Therefore $\psi$ is an
intertwining operator between induced representations of $S$ in
$U(NP)$ and  in $SO(\dim S)$. To f\/ind the surviving fermions, as in
the commutative case \cite{Kapetanakis:1992hf}, we decompose the
adjoint rep.\ of $U(NP)$ under $S_{U(NP)}\times K$,
\begin{gather*}
U(NP) \supset S_{U(NP)} \times K, \\
{\rm adj} \,U(NP) =
\sum_{i} (s_i, k_i).
\end{gather*}
We also decompose the spinor rep. $\sigma$ of $SO(\dim S)$ under $S$
\begin{gather*}
SO(\dim S)  \supset  S ,\\
 \sigma  =  \sum_{e} \sigma_{e}.
\end{gather*}
Then, when we have two identical irreps.\ $s_i = \sigma_e$, there
is a $k_i$  multiplet of fermions surviving in four dimensions,
i.e.\ four-dimensional spinors $\psi(x)$ belonging to the~$k_i$
representation of $K$.

An important point that we would like to stress here is the question
of the renormalizability of the gauge theory def\/ined on $M_4 \times
(S/R)_F$. First we notice that the theory exhibits certain features
so similar to a higher-dimensional gauge theory def\/ined on $M_4
\times  S/R$ that naturally it could be considered as a
higher-dimensional theory too. For instance the isometries of the
spaces $M_4 \times S/R$ and $M_4 \times (S/R)_F$ are the same. It
does not matter if the compact space is fuzzy or not. For example in
the case of the fuzzy sphere, i.e.\ $M_4 \times S^2_N$, the
isometries are $SO(3,1) \times SO(3)$ as in the case of the
continuous space, $M_4 \times S^2$. Similarly the coupling of a~gauge theory def\/ined on $M_4 \times S/R$ and on $M_4 \times (S/R)_F$
are both dimensionful and have exactly the same dimensionality. On
the other hand the f\/irst theory is clearly non-renormalizable, while
the latter is renormalizable (in the sense that divergencies can be
removed by a f\/inite number of counterterms). So from this point of
view one f\/inds a partial justif\/ication of the old hopes for
considering quantum f\/ield theories on non-commutative structures. If
this observation can lead  to f\/inite theories too, it remains as an
open question.

\subsection{The problem of chirality in fuzzy CSDR}\label{section2.4}

Among the great successes of the ordinary CSDR is the possibility to accommodate chiral fermions in the four-dimensional theory~\cite{Manton:1981es}. Needless to say that the requirement of chirality for the four-dimensional fermions is necessary in order for a theory to have a chance to become realistic.

Let us recall the necessary conditions for accommodating chiral fermions in four dimensions when a higher-dimensional gauge theory with gauge group $G$ is reduced over a $d$-dimensional coset space $S/R$ using the CSDR scheme. As we discussed previously, solving the CSDR constraints for the fermion f\/ields
leads to the result that in order to obtain the representations of the four-dimensional unbroken gauge group $H$ under which the
four-dimensional fermions transform, we have to decompose the
representation $F$ of the initial gauge group in which the fermions
are assigned under $R \times H$, i.e.
\[
F= \sum (t_{i},h_{i}),
\]
and the spinor of the tangent space group $SO(d)$ under $R$
\[
\sigma_{d} = \sum \sigma_{j}.
\]
Then for each pair $t_{i}$ and $\sigma_{i}$, where $t_{i}$ and
$\sigma_{i}$ are identical irreducible representations there is an
$h_{i}$ multiplet of spinor f\/ields in the four-dimensional theory.

\looseness=1
In order to obtain chiral fermions in four dimensions we need some
further requirements. The representation of interest, for our
purposes, of the spin group is the spinor representation. This has
dimensions $2^{d\over 2}$ and $2^{(d-1)\over 2}$ for $d$ even and odd
respectively. For odd $d$ the representation is irreducible but for even $d$ it is reducible into two irreducible components of equal
dimension. This splitting exactly gives the possibility to def\/ine
Weyl spinors and to construct a chirality operator. Thus if we are in
odd number of dimensions (where the chirality operator does not
exist) there is no way to obtain chiral fermions. For this reason we
focus only on even dimensions.

The f\/irst possibility is to start with Dirac fermions in $D$ (even)
dimensions. Here we can def\/ine the standard chirality operator
\[ \Gamma^{D+1}=i^{D(D-1)\over 2}\Gamma^1\Gamma^2\cdots \Gamma^D,
\]
with $(\Gamma^{D+1})^2=1$ and $\{\Gamma^{D+1},\Gamma^A\}=0$, where
$\Gamma^A, A=1,\dots,D $ span the Clif\/ford algebra in $D$ dimensions.
This operator has eigenvalues $\pm1$ and distinguishes left and
right spinors. So, it is possible to def\/ine a Weyl basis, where the
chirality operator is diagonal, namely
\[
\Gamma^{D+1}\psi_{\pm}=\pm\psi_{\pm}.
\]
As we mentioned above, in this case $SO(1,D-1)$ has two independent
irreducible spinor representations, $\sigma_D$ and $\sigma'_D$,
under which the Weyl spinors $\psi_+$ and $\psi_-$ transform
respectively. The following branching rule for the spinors holds\footnote{Here the usual notation for two-component Weyl spinors of the Lorentz group $SO(1,3)$ is adopted, namely $\psi_+\rightarrow(1,2)$ and $\psi_-\rightarrow(2,1)$.}
\begin{gather*}
 SO(1,D-1) \supset SO(1,3)\times SO(d), \\
        \sigma_D = (2,1;\sigma_d)+(1,2;\sigma'_d),  \\
        \sigma'_D = (2,1;\sigma'_d)+(1,2;\sigma_d).
\end{gather*}
Then, since we started with a Dirac spinor $\psi=\psi_+ \oplus
\psi_-$ transforming under a representation $F$ of the original gauge
group $G$, following the rule which was stated above it is obvious
that we obtain fermions in four dimensions appearing in equal
numbers of left and right representations of the unbroken gauge
group $H$. Thus, starting with Dirac fermions does not render the
fermions of the four-dimensional theory chiral.

In order to overcome this problem we can make a further restriction and start with Weyl fermions, namely to impose the Weyl condition in
higher dimensions. Then, only one of the $\sigma_D$ and $\sigma'_D$
representations is selected. There are still two cases to
investigate, the total number of dimensions being $4n$ or
$4n+2$. Since we are interested in vacuum conf\/igurations of the form
$M^4\times S/R$ the dimensionality of the internal (coset) space is then
of the form $4n$ or $4n+2$ respectively.

For $D=4n$ ($d=4(n-1)$), the two spinor representations of $SO(d)$
are self-conjugate, meaning that in the decomposition
\begin{gather*} SO(d) \supset  R, \\
        \sigma_{d}  = \sum \sigma_{i},
\end{gather*}
$\sigma_{i}$ is either a real representation or it appears
together with its conjugate representation $\bar{\sigma}_i$. Thus we
are led to consider that the representation~$F$ of~$G$ where the
fermions are assigned has to be complex.
Two important things to note is that $R$ is also required to admit complex
representations (otherwise the decompositions of $\sigma_d$ and
$\sigma'_d$ will be the same, leading to a non-chiral theory) and
that ${\rm rank}\, S= {\rm rank}\, R$ (otherwise~$\sigma_d$ and~$\sigma'_d$ will again
be the same). These requirements still hold in the following
case.

In the case $D=4n+2$ $(d=4(n-1)+2)$, the two spinor representations
of $SO(d)$ are not self-conjugate anymore and
$\sigma'_d=\bar{\sigma}_d$. Now, the decomposition reads as
\begin{gather*} SO(d) \supset  R , \\
        \sigma_{d}  =  \sum \sigma_{i},\\
            \bar{\sigma}_{d}  =  \sum \bar{\sigma}_{i},
\end{gather*}
so we can let $F$ be a vectorlike representation. Then, in the
decomposition
\begin{gather*}
G \supset   R_G\times H, \\
     F =  \sum (t_{i},h_{i}),
\end{gather*}
each term $(t_{i},h_{i})$ will either be self-conjugate or it will
appear with the term $(\bar{t}_{i},\bar{h}_{i})$. According to the
established rule, $\sigma_d$ will provide a left-handed fermion
multiplet transforming under the four-dimensional gauge group as
$f_L=\sum h_i^L$; $\bar{\sigma}_d$ will provide a right-handed
fermion multiplet transforming as $f_R=\sum \bar{h}_i^R$. Since
$h_i^L\sim \bar{h}_i^R$ we are led to two Weyl fermions with the
same chirality in the same representation of the unbroken gauge
group $H$. This is of course a chiral theory, which is the desired result. Moreover, the doubling of the fermions can
be eliminated by imposing the Majorana condition, if applicable\footnote{Let us remind that the Majorana condition can be imposed when the number of dimensions is $D=2,3,8n+4$.}.

Let us use the same spirit in order to investigate the possibility of obtaining chiral fermions in the fuzzy case as well.
We discussed previously that we have to consider the embedding \[ S \subset SO(\dim S), \] concerning fermions in the adjoint.
In order to determine the
surviving fermions, as in the commutative case, we decompose the adjoint rep. of $U(N)$ under
$S_{U(N)}\times K$,
\begin{gather*}
U(N)  \supset  S_{U(N)} \times K, \\ {\rm adj}\, U(N)  =  \sum_{i}
(s_i, k_i) .
\end{gather*}
We also decompose the spinor rep.~$\sigma$ of $SO(\dim S)$ under $S$
\begin{gather*}
SO(\dim S) \supset S, \\
 \sigma  =  \sum_{e} \sigma_{e} .
\end{gather*}
Then, when we have two identical irreps. $s_i = \sigma_e$, there is
a $k_i$  multiplet of fermions surviving in four dimensions, i.e.\
four-dimensional spinors $\psi(x)$ belonging in the $k_i$
representation of~$K$.

Concerning the issue of chirality, the situation is now dif\/ferent.
The main dif\/ference is obviously the modif\/ication
of the rule for the surviving fermions. In the continuous case we
had to embed $R$ in~$SO(d)$, while now the suitable embedding is that of
$S$ in $SO(\dim S)$. Exploring chirality in the continuous case, we had to deal with the
representations of $SO(d)$. Recall that we required $d$ to be even so
that there are two independent spinor representations; therefore in the fuzzy case we require $\dim S$ to be
even.
Moreover, when $d=4n$ we concluded that the representation $F$, where
the fermions are initially assigned, has to be complex. Since in the
fuzzy case we assign the fermions in the adjoint
representation, the case
$\dim S=4n$ would lead to a non-chiral theory. Finally, the case $\dim S=4n+2$
is the only promising one and one would expect to obtain chiral
fermions, as in the continuous case when $d=4n+2$. However, we also
need the further requirement that $S$ admits complex representations,
again in analogy with $R$ admitting complex representations in the
continuous case.

In summary, in order to have a chance to obtain chiral fermions in the case of fuzzy
extra dimensions the necessary requirements are:
\begin{itemize}\itemsep=0pt
\item $\dim S = 4n+2$,
\item $S$ admits complex irreps.
\end{itemize}
The above requirements are quite restrictive; for example they are not satisf\/ied in the case of a~ single fuzzy sphere. In general, using elementary number theory one can show that they cannot be satisf\/ied for any~$S$ being a $SU(n)$, $SO(n)$ or $Sp(n)$ group. Therefore only products of fuzzy spaces have a chance to lead to chiral fermions after dimensional reduction without further requirements. The simplest case which satisf\/ies these requirements is that of a product of two fuzzy spheres, which will be discussed in Section~\ref{section3.4} in the context of dynamical generation of fuzzy extra dimensions.

In conclusion it is worth making the following remark. As we saw above, a major dif\/ference between fuzzy and ordinary CSDR is that in the
fuzzy case one always embeds $S$ in the gauge group $G$ instead of
embedding just $R$ in $G$. A generic feature of the ordinary CSDR in the special case when $S$ is embedded in $G$ is that the fermions in the f\/inal theory are massive \cite{Barnes:1986ea}. According to the discussion in Section~\ref{section2.3.3} the situation in the fuzzy case is very similar to the one we just described. In
fuzzy CSDR the spontaneous symmetry breaking mechanism takes already
place by solving the fuzzy CSDR constraints. Therefore  in the Yukawa sector of the theory we have the results of the spontaneous symmetry breaking,
i.e.\ massive fermions and Yukawa interactions among fermions and the
physical Higgs f\/ield. We shall revisit the problem of chirality in the following section and f\/inally, in Section~\ref{section4}, we shall describe a way to overcome it and obtain chiral four-dimensional theories.

\section{Dynamical generation of fuzzy extra dimensions}\label{section3}

Let us now discuss a further development \cite{Aschieri:2006uw} of
these ideas,
 which addresses in detail the questions of
 quantization and renormalization. This leads to a slightly
modif\/ied model with an extra term in the potential, which
dynamically selects a unique (nontrivial) vacuum out of the many
possible CSDR solutions, and moreover generates a magnetic f\/lux on
the fuzzy sphere. It also allows to show that the full tower of
Kaluza--Klein modes is generated on $S^2_N$. Moreover, upon including fermions, the model of\/fers the possibility of a detailed study of the fermionic sector~\cite{Steinacker:2007ay}. Such a study reveals the dif\/f\/iculty in obtaining chiral low-energy models but at the same time it paves the way out of this problem. Indeed, we shall see in the following section that using orbifold techniques it is possible to construct chiral models in the framework of dynamically generated fuzzy extra dimensions.

\subsection{The four dimensional action}\label{section3.1}

We start with a $SU(N)$ gauge theory on four dimensional Minkowski
space $M^4$ with coordinates~$y^\mu$, $\mu = 0,1,2,3$.  The action
under consideration is
\begin{gather*}
{\cal S}_{YM}= \int d^{4}y\, Tr\,\left( \frac{1}{4g^{2}}\, F_{\mu
\nu}^\dagger F_{\mu \nu} + (D_{\mu}\phi_{{a}})^\dagger
D_{\mu}\phi_{{a}}\right) - V(\phi), 
\end{gather*}
where $A_\mu$ are $SU(N)$-valued gauge f\/ields, $D_\mu =
\partial_\mu + [A_\mu,\cdot]$, and
\[ \phi_{{a}} = - \phi_{{a}}^\dagger, \qquad a=1,2,3 \] are three
antihermitian scalars in the adjoint of $SU(N)$,
\[ \phi_{{a}} \to
U^\dagger \phi_{{a}} U,
\] where $U = U(y) \in SU(N)$. Furthermore,
the $\phi_a$ transform as vectors of an additional global $SO(3)$
symmetry. The potential $V(\phi)$ is taken to be the most general
renormalizable action invariant under the above symmetries, which is
\begin{gather}
V(\phi)   = {\rm Tr}\, \left( g_1 \phi_a\phi_a \phi_b\phi_b +
g_2\phi_a\phi_b\phi_a \phi_b - g_3 \varepsilon_{a b c} \phi_a \phi_b
\phi_c + g_4\phi_a \phi_a \right) \nonumber\\
\phantom{V(\phi)   =}{}  + \frac{g_5}{N}
\,{\rm Tr}\, (\phi_a \phi_a)\,{\rm Tr}\, (\phi_b \phi_b) + \frac{g_6}{N} \,{\rm Tr} (\phi_a
\phi_b)\,{\rm Tr}\, (\phi_a \phi_b) +g_7. \label{pot}
\end{gather}
This may not look very transparent at f\/irst sight, however it can be
written in a very intuitive way. First, we make the scalars
dimensionless by rescaling \[ \phi'_a = R  \phi_a, \] where $R$ has
dimension of length; we will usually suppress $R$ since it can
immediately be reinserted, and drop the prime from now on.  Now
observe that for a suitable choice of $R$,
\[
R = \frac{2 g_2}{g_3},
 \] the potential can be rewritten as
\begin{gather*}
 V(\phi)= {\rm Tr} \left( a^2
(\phi_a\phi_a + \tilde b  \one)^2 + c +\frac{1}{\tilde g^2}
F_{ab}^\dagger F_{ab}  \right) + \frac{h}{N}  g_{ab} g_{ab}
\end{gather*}
for suitable constants $a$, $b$, $c$,~$\tilde g$,~$h$, where
\begin{gather*}
F_{{a}{b}}  =  [\phi_{{a}}, \phi_{{b}}] -
\varepsilon_{abc} \phi_{{c}}  = \varepsilon_{abc} F_c , \qquad
\tilde b  =  b + \frac{d}{N} \, {\rm Tr} \,(\phi_a \phi_a), \qquad
g_{ab}  = {\rm  Tr}(\phi_a \phi_b). 
\end{gather*}
We will omit $c$ from now.
Notice that two couplings were reabsorbed in the def\/initions of~$R$ and~$\tilde b$.
 The potential is clearly positive
def\/inite provided \[ a^2 = g_1+g_2 >0, \qquad \frac 2{\tilde g^2} =
- g_2 >0, \qquad h \geq 0, \] which we assume from now on.  Here
$\tilde b = \tilde b(y)$ is a scalar, $g_{ab} = g_{ab}(y)$ is a
symmetric tensor under the global $SO(3)$, and $F_{ab}=F_{ab}(y)$ is
a $su(N)$-valued antisymmetric tensor f\/ield which will be
interpreted as f\/ield strength in some dynamically generated extra
dimensions below.  In this form, $V(\phi)$ looks like the action of
Yang--Mills gauge theory on a fuzzy sphere in the matrix formulation
\cite{Steinacker:2003sd,Steinacker:2004yu,Carow-Watamura:1998jn,
Presnajder:2003ak}. It dif\/fers from the potential in~(\ref{pot1}) only by the presence of the f\/irst term $a^2
(\phi_a\phi_a + \tilde b)^2$, which is strongly suggested
 by renormalization.
In fact it is necessary for the interpretation as pure YM action,
and we will see that it is very welcome on physical grounds since it
dynamically determines and stabilizes a vacuum, which can be
interpreted as extra-dimensional fuzzy sphere. In particular, it
removes unwanted f\/lat directions.

\subsection{Emergence of extra dimensions and the fuzzy sphere}\label{section3.2}

The vacuum of the above model is given by the minimum of the potential~(\ref{pot}). Finding the minimum of the potential is a rather
nontrivial task, and the answer depends crucially on the parameters in
the potential \cite{Aschieri:2006uw}. The conditions for the global
minimum imply that $\phi_a$ is a~representation of $SU(2)$, with Casimir
$\tilde b$ (where it was assumed for simplicity $ h = 0$).  Then, it
is easy to write down a large class of solutions to the minimum of the
potential, by noting that any decomposition of $N = n_1 N_1 + \cdots +
n_h N_h$ into irreps of $SU(2)$ with multiplicities $n_i$ leads to a
block-diagonal solution
\begin{gather}
 \phi_a = {\rm diag}\,\big(\a_1\, X_a^{(N_1)}, \dots,
\a_k\, X_a^{(N_k)}\big)
\label{solution-general}
\end{gather}
of the vacuum equations, where $\a_i$ are suitable
constants which will be determined below.

It turns out~\cite{Aschieri:2006uw} that
there are essentially only 2 types of vacua:
\begin{enumerate}\itemsep=0pt
\item {\em Type I vacuum:}
 It is plausible that the solution (\ref{solution-general}) with minimal
potential contains only representations whose Casimirs are close to
$\tilde b$. In particular, let $M$ be the dimension of the irrep whose
Casimir $C_2(M)\approx \tilde b$ is closest to $\tilde b$. If
furthermore the dimensions match as $N = M n$, we expect that the
vacuum is given by $n$ copies of the irrep $(M)$, which can be written
as
$\phi_a = \alpha  X_a^{(M)} \otimes \one_{n}$ with low-energy
gauge group $SU(n)$.
\item {\em Type II vacuum:}
Consider again a solution (\ref{solution-general}) with $n_i$ blocks of
size $N_i = \tilde N +m_i$, where $\tilde N$ is def\/ined by
$\tilde b = \frac 14 (\tilde N^2-1)$, and assume that $\tilde N$ is large and
$\frac{m_i}{\tilde N} \ll 1$.  The action
is then given by
\[
 V(\phi) = {\rm Tr} \left( \frac {1}{2\tilde g^2}
\sum_i n_i  m_i^2   \one_{N_i}   + O\left(\frac 1{N_i}\right) \right) \approx
\frac{1}{2\tilde{g}^{2}}  \frac{N}{k}  \sum_{i} n_i   m_{i}^{2} ,
\]
 where $k=\sum n_i$ is the total number of irreps, and the solution
can be interpreted in terms of ``instantons'' (non-Abelian monopoles)
on the internal fuzzy sphere \cite{Steinacker:2003sd}.  Hence in order
to determine the solution of type~(\ref{solution-general}) with minimal
action, we simply have to minimize $\sum_i n_i   m_i^2$, where the $m_i
\in \Z -\tilde N$ satisfy the constraint $\sum n_i  m_i = N - k \tilde
N$.  In this case the solution with minimal potential among
all possible partitions (\ref{solution-general}) is given by
\[
 \phi_a = \left(\begin{array}{cc}\alpha_1\, X_a^{(N_1)}\otimes
\one_{n_1} & 0 \\ 0 & \alpha_2\,X_a^{(N_2)}\otimes \one_{n_2}
             \end{array}\right),
\]
with low-energy gauge group $SU(n_1)\times  SU(n_2) \times U(1)$.
\end{enumerate}

Again, the $X_a^{(N)}$ are interpreted as coordinate functions of a
fuzzy sphere $S^2_{N}$, and the ``scalar'' action
\[ S_{\phi} = {\rm Tr}\, V(\phi) = {\rm Tr}\left(a^2 (\phi_a\phi_a + \tilde b)^2 +
\frac 1{\tilde g^2}  F_{ab}^\dagger F_{ab}\right) \label{S-YM2}\] for
$N \times N$ matrices $\phi_a$ is precisely the action for a $U(n)$
Yang--Mills theory on $S^2_{N}$ with coupling~$\tilde g$, as shown in~\cite{Steinacker:2003sd}. In fact, the new term $(\phi_a\phi_a +
\tilde b)^2$ is essential for this interpretation, since it
stabilizes the vacuum $\phi_a = X_a^{(N)}$ and gives a large mass to
the extra ``radial'' scalar f\/ield which otherwise arises.  The
f\/luctuations of $\phi_a = X_a^{(N)} + A_a$ then provide the
components $A_a$ of a~higher-dimensional gauge f\/ield $A_M = (A_\mu,
A_a)$, and the action can be interpreted as YM theory on the
6-dimensional space $M^4 \times S^2_{N}$, with gauge group depending
on the particular vacuum.  We therefore interpret the vacuum as
describing dynamically generated extra dimensions in the form of a
fuzzy sphere $S^2_{N}$. This geometrical interpretation can be fully
justif\/ied
 by working out the spectrum of Kaluza--Klein
modes.  The ef\/fective low-energy theory is then given by the zero
modes on $S^2_{N}$. This approach provides a clear dynamical
selection of the geometry due to the term $(\phi_a\phi_a + \tilde
b)^2$ in the action.

Perhaps the most remarkable aspect of this model is that the geometric
interpretation and the corresponding low-energy degrees of freedom
depend in a nontrivial way on the parameters of the model, which are
running under the RG group. Therefore the massless degrees of freedom
and their geometrical interpretation depend on the energy scale. In
particular, the low-energy gauge group generically turns out to be
$SU(n_1) \times SU(n_2) \times U(1)$ or $SU(n)$, while gauge groups which are
products of more than two simple components (apart from $U(1)$) do not
seem to occur. The values of $n_1$ and $n_2$ are determined
dynamically, and with the appropriate choice of parameters it is
possible to construct vacuum solutions where they are as small,  such
as~2 and~3~\cite{Aschieri:2006uw}.

It is interesting to examine  the running of the coupling constants
under the RG. $R$ turns out to run only logarithmically, implies
that the scale of the internal spheres is only mildly af\/fected by
the RG f\/low. However, $\tilde b$ is running essentially
quadratically, hence is generically large. This is quite welcome
here: starting with some large $N$, $\tilde{b} \approx
C_2(\tilde{N})$ must indeed be large in order to lead to the
geometric interpretation discussed above. Hence the problems of
naturalness or f\/ine-tuning appear to be rather mild here.

A somewhat similar model has been studied  in
\cite{Andrews:2005cv}, which realizes deconstruction
and a~``twisted'' compactif\/ication of an extra fuzzy sphere based on
a supersymmetric gauge theory. Our model is dif\/ferent and does not
require supersymmetry, leading to a much richer pattern of symmetry
breaking and ef\/fective geometry.

The dynamical formation of fuzzy spaces found here is also related
to recent work studying the emergence of stable submanifolds in
modif\/ied IIB matrix models. In particular, previous studies based on
actions for fuzzy gauge theory dif\/ferent from ours generically only
gave results corresponding to $U(1)$ or $U(\infty)$ gauge groups,
see e.g.~\cite{Azuma:2004ie,Azuma:2005bj,Azuma:2004zq} and
references therein. The dynamical generation of a nontrivial index
on noncommutative spaces has also been observed in~\cite{Aoki:2004sd,Aoki:2006zi} for dif\/ferent models.

Our mechanism may also be very interesting in the context of the
recent observation \cite{Abel:2005rh} that extra dimensions are very
desirable for the application of noncommutative f\/ield theory to
particle physics. Other related recent work discussing the
implications of the higher-dimensional point of view on symmetry
breaking and Higgs masses can be found in
\cite{Lim:2006bx,Dvali:2001qr,Antoniadis:2002ns,Scrucca:2003ra}.
These issues could now be discussed within a renormalizable
framework.

\subsection{Inclusion of fermions}\label{section3.3}

So far we have only discussed the emergence of fuzzy extra dimensions in a four-dimensional and renormalizable pure Yang--Mills theory. Let us now include fermions. First we discuss the Dirac operator on the fuzzy sphere and its spectrum in the type I and type II vacua. Subsequently, we construct a model which dynamically develops fuzzy extra dimensions with the geometry of a product of two fuzzy spheres and study the zero modes of the corresponding Dirac operator in such vacua.

\subsubsection[Fermions on $M^4 \times S^2$ and $M^4 \times S^2_N$]{Fermions on $\boldsymbol{M^4 \times S^2}$ and $\boldsymbol{M^4 \times S^2_N}$}\label{section3.3.1}

We f\/irst recall the classical description of
fermions on $M^4 \times S^2$, formulated in a way which will
generalize to the fuzzy case. This is done using the embedding
$S^2 \subset \R^3$ based on the 7-dimensional Clif\/ford algebra
\begin{gather*}
\Gamma^A = (\Gamma^{\mu},\Gamma^a) = (\one \otimes \gamma^\mu,
\sigma^a \otimes i\gamma_5) .
\end{gather*}
Here $\sigma^a$, $a=1,2,3$ generate the $SO(3)\sim SU(2)$
Clif\/ford algebra.
The $\Gamma^A$ act on $\C^2 \otimes \C^4$  and satisfy
$(\Gamma^A)^\dagger = \eta^{AB} \Gamma^B$ where
$\eta^{AB} = (1,-1,\dots,-1)$ is the 7-dimensional Minkowski metric.
The corresponding 8-component spinors describe
Dirac fermions on $M^4 \times S^2$, and can be viewed as
Dirac spinors on $M^4$ tensored with
2-dimensional Dirac spinors on $S^2\subset\R^3$.
We can def\/ine a 2-dimensional chirality operator
$\chi$ locally at each point of the unit sphere $S^2$ by setting
\[
\chi = x_a \sigma^a ,
\]
which has eigenvalues $\pm 1$. At the north pole
$x_a = (0,0,1)$ of $S^2$ this coincides with $\chi = -i \sigma^1
\sigma^2 = \sigma^3$,
as expected.
The action for a Dirac fermion on $M^4 \times S^2$  can then be written
as
\[
S_{6D} = \int_{M^4} d^4 y  \int_{S^2} d\Omega\, \obar\Psi_D \left(
i \gamma^\mu \partial_\mu
+ i\gamma_5  \not \!\!D_{(2)} + m \right)\Psi_D ,
\]
where
\begin{gather}
\not \!\! D_{(2)}\Psi_D = (\sigma_a L_a  + 1) \Psi_D
\label{2Ddirac-class}
\end{gather}
is the Dirac operator on $S^2$ in ``global'' notation and~$L_a$ are the angular momentum
operators,
while the constant 1 in \eqref{2Ddirac-class}  ensures
$\{\not \!\!\! D_{(2)},\chi\}=0$ and ref\/lects the curvature of~$S^2$.
This is
equivalent to the standard formulation in terms of a comoving frame,
but more appropriate for the fuzzy case.

As we discussed in Section~\ref{section2.4}, chiral (Weyl) spinors $\Psi_\pm$ on  $M^4 \times S^2$ are then def\/ined  using the
6D chirality operator
\[
\Gamma =  \gamma_5 \chi ,
\]
where $\g_5$ is the four-dimensional chirality operator, and they satisfy $\Gamma \Psi_\pm = \pm \Psi_\pm$.
They contain both chiralities from the four-dimensional point of view,
\[
\Psi_\pm = (0,1;\pm) + (1,0;\mp) ,
\]
where $(0,1;\pm)$ denotes a Weyl spinor $\psi_\a$ on $M^4$ with
eigenvalue $\pm 1$ of $\chi$, and $(0,1;\mp)$  a dotted Weyl spinor
$\obar\psi^{\dot\a}$ on $M^4$ with eigenvalue $\mp 1$ of $\chi$.
These components are of course mixed under the six-dimensional rotations.

Let us now collect the main facts about the
``standard'' Dirac operator on the fuzzy sphere~\cite{Grosse:1994ed},
which is given by the following analog of \eqref{2Ddirac-class}
\[
\not \!\! D_{(2)}\Psi = \sigma_a [i X_a,\Psi] +  \Psi ,
\]
where $[X_a,X_b] = \varepsilon_{abc} X_c$ generate the fuzzy sphere
as explained before. Let us recall that
$X_a$ is antihermitian here. $\not \!\! D_{(2)}$ acts on 2-component spinors
\[
\Psi = \left(\begin{array}{c} \psi_{1} \\
               \psi_{2} \end{array}\right).
\]
For spinors in the adjoint of the gauge group, the generators $X_a$
are replaced by the covariant coordinates $\phi_a$, and
the gauged Dirac operator is
\[
\not \!\! D_{(2)}\Psi = \sigma_a [i\phi_a,\Psi] +  \Psi.
\]
Let us note that there does not exist a chirality operator which anticommutes
with $\not \!\! D_{(2)}$ and has
eigenvalues $\pm 1$; this follows from the spectrum of $\not \!\!
D_{(2)}$, which will be determined below.

\subsubsection[The spectrum of $\not \!\! D_{(2)}$ in a type I vacuum]{The spectrum of $\boldsymbol{\not \!\! D_{(2)}}$ in a type I vacuum}

Since $\not \!\! D_{(2)}$ commutes with the $SU(2)$ group of rotations,
the eigenmodes of  $\not \!\! D_{(2)}$ in the type I vacuum
are obtained by decomposing the spinors into irreps of $SU(2)$
\begin{gather}
\Psi  \in   (2)  \otimes (N) \otimes (N)
 = (2)\tens ((1)\oplus(3) \oplus \cdots \oplus (2N-1))\nonumber\\
 \phantom{\Psi}{} =  ((2) \oplus (4) \oplus \cdots  \oplus(2N))
   \oplus ((2) \oplus\cdots \oplus (2N-2))
 =:  ( \Psi_{+,(n)}
        \oplus   \Psi_{-,(n)}) .
\label{spinor-decomp}
\end{gather}
This def\/ines the spinor harmonics $\Psi_{\pm,(n)}$ which live in
the $n$-dimensional representation of $SU(2)$ denoted by $(n)$ for
$n=2,4,\dots,2N$, excluding $\Psi_{-,(2N)}$. The eigenvalue
of $\not \!\! D_{(2)}$ acting on these states can be determined easily
using some $SU(2)$ algebra \cite{Steinacker:2003sd}:
\[
\not \!\! D_{(2)}\Psi_{\pm,(n)}
= E_{\delta = \pm,(n)} \Psi_{\pm,(n)} ,
\]
where
\[
E_{\delta = \pm,(n)}\, \approx\, \frac{\a}2\, \left\{\begin{array}{lll}
n, & \delta = 1, & n=2,4,\dots, 2N ,\\
-n, & \delta = -1,&  n=2,4,\dots, 2N-2
\end{array}\right.
\]
assuming $\a \approx 1$; this is exact for $\a=1$.
We note that with the
exception of $\Psi_{+,(2N)}$, all eigenstates come in pairs
$(\Psi_{+,(n)},\Psi_{-,(n)})$ for $n=2,4,\dots, 2N-2$, which have opposite
eigenvalues $\pm \frac{\a}2 n$ of $\not \!\! D_{(2)}$.

\subsubsection[The spectrum of $\not \!\! D_{(2)}$ in a type II vacuum]{The spectrum of $\boldsymbol{\not \!\! D_{(2)}}$ in a type II vacuum}

Consider now a  type II vacuum,
\[
\left(\begin{array}{cc} \a_1 X_a^{N_1}\otimes \one_{n_1} & 0 \\
            0 & \a_2 X_a^{N_2}\otimes \one_{n_2}
\end{array}\right).
\]
We decompose the spinors according to this block-structure as
\[
\Psi =
\left(\begin{array}{cc} \Psi^{11} & \Psi^{12}\\ \Psi^{21} & \Psi^{22}
\end{array}\right).
\]  The analysis for
the diagonal blocks is the same as before, and they describe fermions
in the adjoint of $SU(n_1)$ resp.\ $SU(n_2)$.  The of\/f-diagonal blocks
however describe fermions in the
bifundamental $(n_1) \times (\obar n_2)$ of $SU(n_1) \times
SU(n_2)$, and those will provide the interesting low-energy sector.
For the moment we ignore the extra
$SU(n_i)$ structure. Assuming $N_1 \neq N_2$,
their decomposition \eqref{spinor-decomp} into irreps of
the global $SU(2)$ now reads
\begin{gather*}
\Psi^{12}  \in   (2)  \otimes (N_1) \otimes (N_2)
 = (2) \tens ((1+|N_2\!-N_1|)\oplus(3+|N_2\!-N_1|) \oplus \cdots \oplus (N_1\!+N_2\!-1)) \\
\phantom{\Psi^{12}}{} =  ((|N_2-N_1|+2) \oplus (|N_2-N_1|+4) \oplus \cdots  \oplus(N_1+N_2))   \\
\phantom{\Psi^{12}  \in}{}   \oplus ((|N_2-N_1|) \oplus (|N_2-N_1|+2) \oplus \cdots
            \oplus (N_1+N_2-2))  =: \big(\Psi^{12}_{+,(n)}  \oplus \Psi^{12}_{-,(n)}\big)
\end{gather*}
def\/ining the spinor harmonics $\Psi^{12}_{\pm,(n)}$ which live in
the representation $(n)$ of $SU(2)$. A similar decomposition
holds for $\Psi^{21} \in  (2)  \otimes (N_2) \otimes (N_1)$.

\subsection[Dynamical generation of fuzzy $S^2\times S^2$ and mirror fermions]{Dynamical generation of fuzzy $\boldsymbol{S^2\times S^2}$ and mirror fermions}\label{section3.4}

As we have previously discussed in Section~\ref{section2.4} a single fuzzy sphere is not a good candidate in order to obtain chiral fermions in four dimensions. However, a product of two fuzzy spheres~\cite{Behr:2005wp} is more promising, since its isometry group is $S=SU(2)\times SU(2)$ with dimension $\dim S=6$, which is of the form $4n+2$, as required. Therefore we shall study this case in the context of dynamical generation of fuzzy extra dimensions and explore the fermionic sector in a type II vacuum where f\/luxes can be included. It will turn out that the fermions are accommodated in complex, bifundamental representations which however come in pairs of opposite chirality. This picture corresponds then to mirror fermions~\cite{Maalampi:1988va}.

\subsubsection{The action}

Let us consider a $SU(N)$ Yang--Mills gauge theory in four-dimensional
Minkowski spacetime, coupled to six scalars
$\phi_a = \phi_a^\dagger$ ($a=1,\dots, 6$) and
four Majorana spinors $\chi_p$ ($p=1,\dots, 4$) in the adjoint representation
of the $SU(N)$.
Moreover, we assume that the $\phi_a$ transform in the vector representation
of a global $SU(4) \cong SO(6)$ group and
the $\chi_p$ in the fundamental of the~$SU(4)$. The above particle spectrum
coincides with the spectrum of the ${\cal N}=4$ supersymmetric Yang--Mills
theory (SYM) \cite{Brink:1976bc}, where the global $SU(4)$ is the $R$-symmetry of the theory.
The corresponding action is given by
\begin{gather}
{\cal S}_{\rm YM}  =  \int d^4 x\Biggl[ \Tr\left(-\frac 1{4} F_{\mu\nu}F^{\mu\nu}
+ \frac 12 \sum\limits_{a=1}^6\,D^\mu\phi_a D_\mu \phi_a
- V(\phi)\right)  \nonumber\\
 \phantom{{\cal S}_{\rm YM}  =}{}
 +  \frac 12 \Tr \left(i\bar\chi_p \slashed D \chi_p
  + g_4 (\Delta_R^a)_{pq}\,\bar\chi_p R [\phi_a,\chi_q]
  - g_4 (\Delta_L^a)_{pq}\,\bar\chi_p L [\phi_a,\chi_q] \right)\Biggl],
\label{YM-action-4D}
\end{gather}
where the potential has the form
\begin{gather}
\label{4dpot}
 V(\phi) = V_{{\cal N}=4}(\phi)+ V_{\rm break}(\phi).
 \end{gather}
 In (\ref{4dpot})
the f\/irst term corresponds to the potential of the ${\cal N}=4$ SYM
theory, which is explicitly given by
 \begin{gather}
 \label{n4pot} V_{{\cal N}=4}(\phi)=-\frac{1}{4}g_4^2 \sum\limits_{a,b}
  [\phi_a,\phi_b]^2,
  \end{gather}
   while the second term corresponds to an
explicit $R$-symmetry-breaking potential, which breaks the ${\cal N}=4$
supersymmetry as well as the global $SU(4)$ symmetry. We shall see in Section~\ref{section4} that this potential actually corresponds to a set of ${\cal N}=1$ soft supersymmetry breaking terms, which will be explicitly presented later.

In the above expressions $\mu,\nu=0,1,2,3$ are four-dimensional spacetime
indices and $D_\mu = \partial_\mu - i g [A_\mu,\cdot]$ is the four-dimensional
covariant derivative in the adjoint representation. The projection operators
 $L$ and $R$ are, as usual, def\/ined as $L= \frac 12 (\one - \gamma_5)$ and $R=
\frac 12 (\one + \gamma_5)$. The
$(\Delta_{L}^a)_{pq}$ and $(\Delta_{R}^a)_{pq}$ are the intertwiners of
the $\mathbf{4}\times\mathbf{4}
\to \mathbf{6}$ and $\mathbf{\bar 4}\times\mathbf{\bar 4} \to \mathbf{6}$
respectively, namely they are Clebsch--Gordan coef\/f\/icients that couple two
$\mathbf{4}$s to a $\mathbf{6}$. The Yukawa interactions in~(\ref{YM-action-4D})
are separately invariant
under the $SU(4)$, since the~$R \chi_p$ transforms in the $\mathbf{4}$
and the $L\chi_p$ in the~$\mathbf{\bar 4}$ of the~$SU(4)$.

The action \eqref{YM-action-4D} without the term $V_{\rm break}$ can be obtained by a toroidal dimensional reduction of ten-dimensional ${\cal N}=1$ SYM theory~\cite{Brink:1976bc}.
The corresponding ten-dimensional action is
\[
{\cal S}_{D=10} = -\frac 1{4g^2_{10}} \int d^{10} x\, \Tr \, F_{MN} F^{MN}
  +  \frac 12 \int d^{10} x\, \Tr \, \bar\Psi i\Gamma^M  D_M \Psi ,
\]
where
\[
 D_M = \partial_M - ig [A_M,\cdot],
\] capital Latin letters denote ten-dimensional indices, i.e.\ $M= 0, \dots, 9$ and $\Psi$ is a ten-dimensional Majorana--Weyl spinor.
Considering a compactif\/ication of the form $M^4\times Y$, the scalars are obtained
from the internal components of the higher-dimensional gauge f\/ield according to
the splitting
\[
A_M = (A_\mu,\Phi_{3+a}), \qquad a=1,\dots,6.
\]
The ten-dimensional Clif\/ford algebra, generated by
$\Gamma_M$, naturally separates into
a four-di\-men\-sio\-nal and a six-dimensional one as follows,
\[
\Gamma_M = (\Gamma_{\mu},\Gamma_{3+a}), \qquad  \Gamma_{\mu} = \gamma_
{\mu}\otimes\one_8, \qquad \Gamma_{3+a}  =  \gamma_5\otimes\Delta_a.
\]
Here the $\gamma_{\mu}$ def\/ine the four-dimensional Clif\/ford algebra and they are chosen
to be purely imaginary,
corresponding to the Majorana representation in four dimensions (see Appendix~\ref{appendixA}),
while the $\Delta_a$ def\/ine the six-dimensional Euclidean Clif\/ford
algebra and they are chosen to be real and antisymmetric.
Then it is straightforward to see that $\gamma_0 =
\gamma_0^\dagger= - \gamma_0^T$ and
$\gamma_i = -\gamma_i^\dagger=  \gamma_i^T$.
The ten-dimensional chirality operator is
\[
\Gamma^{(11)} = \gamma_5 \otimes\Gamma^{(Y)},
\]
where the four- and six-dimensional chirality operators are def\/ined as
\begin{gather*}
\gamma_5  =  -i\gamma_0 \cdots \gamma_3 = \gamma_5^\dagger
= - \gamma_5^T,  \\
\Gamma^{(Y)}  =  -i\Delta_1 \cdots \Delta_6 = \big(\Gamma^{(Y)}\big)^\dagger
= -\big(\Gamma^{(Y)}\big)^T.
\end{gather*}
Let us denote the ten-dimensional charge conjugation operator as
\[
\cC = C^{(4)} \otimes C^{(6)},
\]
where $C^{(4)}$ is the four-dimensional charge
conjugation operator and $C^{(6)}=\one_8$ in our conventions. This operator
satisf\/ies, as usual, the relation
\[
 \cC\Gamma^M\cC^{-1}=-\big(\Gamma^M\big)^T.
\]
  Then
the Majorana--Weyl condition in ten dimensions is\footnote{Note that $T$ transposes only the spinor.}
\[
\Psi^C = \cC \bar\Psi^T
 {=} \Psi,
 \] where
\[
\bar\Psi  =  \Psi^T \cC^T, \qquad
\Psi^\dagger  =  \Psi^T \cC^T \gamma_0=\Psi^T.
\]
Let us note that in the Majorana representation, where the $\gamma_\mu$
are imaginary, the four-di\-men\-sio\-nal charge conjugation operator is
$C^{(4)} = -\gamma_0$.

Performing a trivial dimensional reduction from ten to four dimensions, i.e.
assuming that all f\/ields do not depend on the internal coordinates, it is
well-known that the Yang--Mills part of the ten-dimensional action leads to
the bosonic part of the ${\cal N}=4$ SYM in four dimensions, as in~(\ref{YM-action-4D}), with the potential term having the form~(\ref{n4pot}).
The couplings $g_4$ and $g_{10}$ are related through the volume $V$ of the
internal six-dimensional torus as $g_4=\frac{g_{10}}{\sqrt{V}}$.

The reduction of the Dirac term is performed similarly. The Majorana--Weyl
spinor $\Psi$
has the form
\begin{gather*}
\Psi  =  \sum_{p=1}^4 \big( R\chi_p \otimes  \eta_p
 + L\chi_p \otimes   \eta_p^* \big) , \qquad
\bar\Psi  = \sum_{p=1}^4 \big(\bar\chi_p L \otimes  \eta_p^\dagger
 + \bar\chi_p R \otimes   \eta_p^T \big),
\end{gather*}
where the $\chi_p$ are four-dimensional Majorana spinors
and the $\eta_p$ are the four complex eigenvectors of the $\Gamma^{(Y)}$ with eigenvalue~$+1$. Since the $\Gamma^{(Y)}$ is purely imaginary the~$\eta_p^*$ have eigenvalue~$-1$.
Assuming that the spinor is independent of the extra-dimensional coordinates,
the dimensional reduction of the Dirac term of the ten-dimensional action leads
in four dimensions to the kinetic term for the spinor $\chi_p$ and the Yukawa
couplings as they appear in~(\ref{YM-action-4D}). In particular, the Yukawa
couplings arise from the term
\[
\Tr\, \bar\Psi  i \slashed D_{(6)}\Psi
= \Tr \,\bar\Psi   \Delta^{a}[\Phi_a,\Psi],
\] where $\slashed D_{(6)}$ denotes the Dirac operator on the internal space,
which satisf\/ies
\[
\big\{\slashed D_{(6)},\Gamma^{(Y)}\big\} =0
\]
 and it will be related in the ensuing to the ef\/fective Dirac operator on the fuzzy
extra dimensions.

\subsubsection[Type I vacuum and fuzzy $S^2 \times S^2$]{Type I vacuum and fuzzy $\boldsymbol{S^2 \times S^2}$}

We now assume that the renormalizable potential in the four-dimensional action
admits vacua corresponding to the product of two
fuzzy spheres, i.e.
\begin{gather}
\phi_i^L  \equiv  \phi_i  =  \a_L  \la_i^{(N_L)}\otimes \one_{N_R}
\otimes\one_{n},\nonumber\\
\phi_{i}^R  \equiv  \phi_{3+i}   =  \a_R  \one_{N_L}\otimes\la_{i}^{(N_R)}
\otimes\one_{n}, \qquad i=1,2,3,
\label{vacuum-typeI-S2S2}
\end{gather}
where $\la_i^{(N)}$ denotes the generator of the
$N$-dimensional irreducible representation of $SU(2)$ and therefore
\begin{gather*}
  [\phi_i^L,\phi_j^L]  =  i \a_L  \varepsilon_{ijk} \phi_k^L,\qquad \phi_i^L
\phi_i^L  =  \a_L^2 \frac{N_L^2-1}4,
\end{gather*}
and similarly for the $\phi_i^R$. Moreover the two algebras commute with each other,
 i.e.
 \[
   [\phi_i^L,\phi_j^R] = 0.
   \]

The vacuum \eqref{vacuum-typeI-S2S2} can be obtained by choosing the potential $V(\Phi)$
to have the following form,
\begin{gather}
V[\Phi] = a_L^2 \big(\phi_i^L\phi_i^L + b_L\one\big)^2
 + a_R^2 \big(\phi_i^R\phi_i^R + b_R\one\big)^2
 + \frac 1{g_L^2} F_{ij}^L F_{ij}^L + \frac 1{g_R^2} F_{ij}^R F_{ij}^R, \label{V-S2S2}
\end{gather}
 where
\[
F_{ij}^L  =  [\phi_i^L,\phi_j^L] - i\varepsilon_{ijk} \phi_k^L, \qquad
F_{ij}^R  =  [\phi_i^R,\phi_j^R] - i\varepsilon_{ijk} \phi_k^R,
\]
 which will be interpreted as f\/ield strengths on the spontaneously generated fuzzy spheres, as in Section~\ref{section3.2}.
The potential (\ref{V-S2S2}) breaks the global $SO(6)$ symmetry down to $SO(3)_L \times SO(3)_R$ and for suitable parameters $a_{L/R}$, $b_{L/R}$, $g_{L/R}$, its stable global minimum is indeed given by \eqref{vacuum-typeI-S2S2} up to
$U(N)$ gauge transformations,
provided that
\begin{gather}
N = N_L N_R  n .
\label{typeI-condition}
\end{gather}
Such a vacuum should be interpreted as a stack of $n$
fuzzy branes with geometry $S^2_L \times S^2_R$ and in the present construction it breaks the
gauge group $SU(N)$ down to $SU(n)$.

\subsubsection[Operators on $S^2_L\times S^2_R$]{Operators on $\boldsymbol{S^2_L\times S^2_R}$}

Having in mind a compactif\/ication on
$S^2_L\times S^2_R \subset \R^6$, we organize the internal
$SO(6)$ structure according to its subgroup
$SO(3)_L \times SO(3)_R$. Then, if $\Delta_a$ def\/ine the six-dimensional Euclidean Clif\/ford
algebra  (see Appendix~\ref{appendixA}) it is natural to adopt the notation
\[
\Delta_i^L = \Delta_i, \qquad \Delta_i^R = \Delta_{3+i},
\qquad i=1,2,3 .
\]
Let us consider the following $SO(3)_L \times SO(3)_R$
invariant operators on each sphere \cite{Behr:2005wp},
\begin{gather*}
\chi_{L} =  \frac i{2R_L}  \Delta_i^L \big\{\phi_i^L,\cdot \big\}
 \sim  \frac i{R_L}  \Delta^i_L x_i^L ,  \qquad
\chi_{L,{\rm tang}}  =  \Gamma^{(Y)}_L \chi_L, \qquad
\Gamma^{(Y)}_{L}  =  \Delta_1 \Delta_2 \Delta_3,
\end{gather*}
where
\[
R_L = \a_L N_L
\]
denotes the radius of $S^2_L$
and the operators $\chi_{R}$, $\chi_{R,{\rm tang}}$ and $\G^{(Y)}_R$ are def\/ined similarly. Here $\sim$ denotes the semi-classical limit, i.e.\ the limit $N\rightarrow\infty$. These operators
are hermitian,
 and they satisfy the relations
\begin{gather*}
\{\chi_{L/R}, \Gamma^{(Y)}\}  =  [\chi_L \chi_R,\Gamma^{(Y)}] = 0, \qquad
\{\chi_L, \chi_R\}  =  0  ,    \\
 [\chi_{L,{\rm tang}}, \chi_{R,{\rm tang}}]  =  0, \qquad
\chi_{L/R}^2 \sim   \one    \sim  \chi_{L/R,{\rm tang}}^2
\end{gather*}
in a  $S^2_L \times S^2_R$ vacuum \eqref{vacuum-typeI-S2S2}.
In order to understand their
meaning, let us consider $S^2_L$.
On the north pole with $x_1\sim0, x_2\sim 0, x_3 \sim R_L$,
the tangential chirality operator is given by
$\chi_{L,{\rm tang}}\sim i\Delta_1 \Delta_2$, while
the operator $\chi_{L} \sim i\Delta_3$ is perpendicular.
Therefore the $SU(2)_L \times SU(2)_R$-invariant operator
\[
\chi^\perp := i \chi_{L}\chi_{R},
\]
which squares to one, $(\chi^\perp)^2 \sim 1$, corresponds to the chirality
operator on the
two-dimensional space which is perpendicular to
$S^2_L\times S^2_R \subset \R^6$. In addition, the operator{\samepage
\[
\chi_{\rm tang}  :=  \Gamma^{(Y)}\chi^\perp
= - \chi_{L,{\rm tang}} \chi_{R,{\rm tang}}
\sim \Delta_1 \Delta_2\Delta_4 \Delta_5, \qquad (\chi_{\rm tang})^\dagger
 = \chi_{\rm tang}
\]
is the tangential chirality operator on $S^2_L \times S^2_R$.}

In order to understand
the fuzzy Kaluza--Klein modes,
it is important to understand the
relation of the Dirac operator on the internal space, let us call it $\slashed D_{(6)}$, with the fuzzy Dirac operators
on~$S^2_L$ and~$S^2_R$. We note that
in the Majorana representation of the six-dimensional Clif\/ford algebra, we have
\[
-i\Gamma^{(Y)}_L \D_i^L = \one_2 \otimes \gamma_L^i,
\]
where $\gamma_L^i = U^{-1} (\sigma_i\otimes \one_2) U$
is essentially a double-degenerate
representation of the three-dimensional Clif\/ford algebra.
This allows to write
\[
\Delta^i_{L} [\phi_i^{L},\Psi]
+ i\a_{L}\Gamma^{(Y)}_{L}\Psi = i\Gamma^{(Y)}_L \slashed D_{S^2_{L}}\Psi,
\]
where $\slashed D_{S^2_{L}}$ is the standard Dirac operator on
the fuzzy $S^2_L$, as described in Section~\ref{section3.3.1}.
Note that one usually works with two-component
spinors on the fuzzy sphere, where the tangential chirality operator
is given by $\sigma_1 \sigma_2 =i \sigma_3$. Here
$\Delta_3$ is independent of $\Delta_1 \Delta_2$ and therefore
$\chi_{L,{\rm tang}}$
is the proper tangential chirality operator
on the $S^2_L$, rather than $\chi_{L}$.
We thus obtain the relation of $\slashed D_{(6)}$
with a ``tangential'' Dirac operator on
$S^2 \times S^2 \subset \R^6$:
\[
\slashed D_{(6)} = \slashed D_{S^2\times S^2}
 -   \a_L \Gamma^{(Y)}_L - \a_R \Gamma^{(Y)}_R,  \]
  where
  \[
\slashed D_{S^2\times S^2}
=    \Gamma^{(Y)}_L\slashed D_{S^2_L}
 + \Gamma^{(Y)}_R\slashed D_{S^2_R}.
\]
Then the term
\[
 \int\bar\Psi i\slashed{D}_{(6)}\Psi,
 \] which gives rise to the
Yukawa terms in the action~(\ref{YM-action-4D}), becomes
\[
 \cS_{yuk}= \int \obar \Psi i \slashed D_{S^2\times S^2} \Psi+\cS_{\rm shift},
 \]
where the shift action
\[
\cS_{\rm shift}= \int i \, {\rm Tr}\,\obar\Psi\gamma_5\big(\a_L \Gamma^{(Y)}_L + \a_R\Gamma^{(Y)}_R\big)
\Psi
\label{S-shift}
\]
is recognized as curvature ef\/fect.
One can show that $\slashed D_{S^2\times S^2}$ reduces
in the semi-classical limit to the Dirac operator on $S^2_L \times S^2_R$
in the above background geometry~\eqref{vacuum-typeI-S2S2}~\cite{Behr:2005wp}.

\subsubsection{Type II vacuum and the zero-modes}

In order to obtain massless fermions, it is necessary to add
magnetic f\/luxes $m_L$, $m_R$ on the two spheres.
As explained previously, this is realized in a slightly modif\/ied class
of vacua, called ``type II vacua''.
In the present case such a vacuum has the form
\begin{gather*}
\phi_i
 =  \left(\begin{array}{cc}\a_1  \la_i^{(N_L^1)}\otimes \one_{N_R^1}
\otimes\one_{n_1} & 0 \\ 0
& \a_2 \la_i^{(N_L^2)}\otimes \one_{N_R^2}\otimes\one_{n_2}
             \end{array}\right),   \\
\phi_{3+i}
 =  \left(\begin{array}{cc}\a_3  \one_{N_L^1}\otimes\la_{i}^{(N_R^1)}
\otimes\one_{n_1} & 0 \\ 0
& \a_4 \one_{N_L^2}\otimes\la_{i}^{(N_R^2)}\otimes\one_{n_2}
             \end{array}\right),\qquad i=1,2,3.
\end{gather*}
The commutant of these generators, i.e.\ the
unbroken gauge group, is
$SU(n_1) \times SU(n_2) \times U(1)_Q$, where the $U(1)_Q$
has generator
\[
Q = \left(\begin{array}{cc} \frac 1{N_R^1 N_L^1 n_1}\one & 0 \\ 0
& -\frac 1{N_R^2 N_L^2 n_2}\one
             \end{array}\right).
\]
This vacuum corresponds to a splitting
\[
 N = n_1 N_L^1 N_R^1 + n_2 N_L^2 N_R^2,
\]
which is more generic than \eqref{typeI-condition}. It
determines a splitting of the
fermionic wavefunction
\[
\Psi = \left(\begin{array}{cc} \Psi^{11} & \Psi^{12} \\
                              \Psi^{21} & \Psi^{22}
\end{array}\right),
\]
where $\Psi^{12}$ transforms in the bifundamental representation $(n_1)\otimes
(\obar n_2)$ of the
$SU(n_1) \times SU(n_2)$ and
$\Psi^{21}$  in the $(\obar n_1)\otimes (n_2)$.
The Majorana condition $\Psi^+\equiv\Psi^{\dagger T} = \Psi$ implies
\[
(\Psi^{11})^{+} =\Psi^{11},\qquad
(\Psi^{22})^{+} =\Psi^{22},\qquad
(\Psi^{12})^{+} =\Psi^{21} .
\]
The interpretation of this vacuum is as a stack of
$n_1$ fuzzy branes
and a stack of $n_2$ fuzzy branes
 with geometry $S^2_L \times S^2_R$.
However, these fuzzy spheres carry
magnetic f\/lux under the unbroken $U(1)_Q$
given by~\cite{Steinacker:2003sd}
\[
m_L= N_L^1-N_L^2, \qquad m_R = N_R^1-N_R^2,
\]
on $S^2_L$ and $S^2_R$ respectively.
Since the fermions $\Psi$ transform in
the adjoint representation, the diagonal components $\Psi^{11}$ and
$\Psi^{22}$ are unaf\/fected, but the
of\/f-diagonal components $\Psi^{12}$ and
$\Psi^{21}$ feel this magnetic f\/lux and develop chiral zero
modes according to the index theorem. This can also be seen very
explicitly in the fuzzy case~\cite{Steinacker:2007ay}.
For example, a f\/lux $m_L> 0$ on  $S^2_L$ implies
that there are (would-be) zero modes $\Psi^{12}_{(m_L)}$
for $\slashed D_{S^2_{L}}$ with $\chi_{L,{\rm tang}} = +1$, and $\Psi^{21}_{(m_R)}$
with $\chi_{L,{\rm tang}} = -1$.

To be specif\/ic, assume that  $m_L>0$ and $m_R>0$.
Then
there exist  (``would-be'', approximate)
zero modes $\Psi^{12}_{(m_L,m_R)}$
of both $\slashed D_{S^2_{L}}$
and $\slashed D_{S^2_{R}}$ and therefore of
$\slashed D_{S^2\times S^2}$,
with def\/inite chirality\footnote{To simplify the notation we assume
that the operators $\chi$, $\slashed D_{S^2}$
are def\/ined appropriately such that
these relations hold exactly. Otherwise
the stated eigenvalues of $\chi$ and $\slashed D_{S^2}$
are approximate up to~$O(\frac 1N)$ corrections.
Since we are mainly interested
in the structure of the would-be zero modes, we do not keep
track of these~$O(\frac 1N)$ corrections here.}
\begin{gather}
\chi_{L,{\rm tang}} \Psi^{12}_{(m_L,m_R)}  =   \Psi^{12}_{(m_L,m_R)}
= \chi_{R,{\rm tang}} \Psi^{12}_{(m_L,m_R)}, \nonumber\\
\chi_{\rm tang} \Psi^{12}_{(m_L,m_R)}   =   -\Psi^{12}_{(m_L,m_R)}.
\label{zeromodes-full-12}
\end{gather}
There are also the ``conjugate'' zero modes
$\Psi^{21}_{(m_L,m_R)}$, which satisfy
\begin{gather}
\chi_{L,{\rm tang}} \Psi^{21}_{(m_L,m_R)}  =  - \Psi^{21}_{(m_L,m_R)}
= \chi_{R,{\rm tang}} \Psi^{21}_{(m_L,m_R)},   \nonumber\\
\chi_{\rm tang} \Psi^{21}_{(m_L,m_R)}   =  - \Psi^{21}_{(m_L,m_R)}.
\label{zeromodes-full-21}
\end{gather}
All the other Kaluza--Klein modes have both chiralities and acquire a mass due to $\slashed D_{S^2\times S^2}$.

Motivated by the properties of the zero modes which are encoded in
(\ref{zeromodes-full-12}) and (\ref{zeromodes-full-21}) let us now def\/ine
the following operators,
\begin{gather}
\Pi_L \Psi := \gamma_5\chi_{L,{\rm tang}}\Psi,  \qquad
\Pi_R \Psi := \gamma_5\chi_{R,{\rm tang}}\Psi,\label{chiral-map}
\end{gather}
which satisfy $\Pi_L^2 \sim \one \sim \Pi_R^2.$
They are clearly compatible with the ten-dimensional
Weyl condition and also with the ten-dimensional Majorana condition
$\Psi^\dagger = \Psi^T$.
Consequently they are well-def\/ined and as we shall exhibit in the following
they will select the chiral sectors of our model.
In order to understand the qualitative structure of the zero modes, in particular their chirality from the four-dimensional point of view, it is enough to consider the semi-classical limit.
On the north pole we have,
\begin{gather*}
\chi_{L,{\rm tang}}  \sim  i\D_1 \D_2 =  \one\otimes \sigma^3 \otimes \sigma^2,\qquad 
\chi_{R,{\rm tang}}  \sim  i\D_4 \D_5 =  \one\otimes \sigma^2 \otimes \sigma^3,\\ 
\chi_{\rm tang}  \sim   \one\otimes \sigma^1 \otimes \sigma^1, \qquad
\chi^\perp  \sim  \sigma_2\otimes \sigma^1 \otimes \sigma^1.
\end{gather*}
Then, the unique solution of  \eqref{zeromodes-full-12} has the form
\begin{gather}
\Psi^{12}_{(m_L,m_R)} \sim\left(\barr{l}\rho^{12}\\ \eta^{12}\earr\right)
\otimes \left(\left(\barr{l}1\\1\earr\right)\otimes\left(\barr{l}1\\1\earr\right)
 -  i\left(\barr{l}1\\-1\earr\right)\otimes\left(\barr{l}1\\-1\earr\right)\right),
\label{zeromodes-spinor-12}
\end{gather}
where $\rho^{12}$, $\eta^{12}$ are four-dimensional Dirac spinors. Similarly,
the unique solution of  \eqref{zeromodes-full-21} has the form
\begin{gather*}
\Psi^{21}_{(m_L,m_R)} \sim\left(\barr{l}\rho^{21}\\ \eta^{21}\earr\right)
\otimes \left(\left(\barr{l}1\\1\earr\right)\otimes\left(\barr{l}1\\1\earr\right)
 +  i\left(\barr{l}1\\-1\earr\right)\otimes\left(\barr{l}1\\-1\earr\right)\right).
\end{gather*}
The Weyl condition $\Gamma^{(11)}\Psi = \Psi$ implies
\begin{gather*}
\gamma_5\Psi   =  \Gamma^{(Y)} \Psi
  =   - (\sigma_2\otimes\one\otimes\one) \Psi,  \qquad
i\eta^{12}  =  \gamma_5\rho^{12},\qquad
i\eta^{21}  =  \gamma_5\rho^{21},
\end{gather*}
so that the would-be zero mode reduces essentially to
\[
\Psi_{(m)}^{12} \sim \left(\barr{l}\rho^{12}\\
  -i \gamma_5 \rho^{12}\earr\right), \qquad
\obar{\Psi_{(m)}^{12}} \sim (\obar{\rho^{12}},
  -i \obar{\rho^{12}}\gamma_5 ),
\]
dropping the remaining tensor factors in \eqref{zeromodes-spinor-12}.
The Majorana condition, $\Psi^+:=\Psi^{\dag T}=\Psi$, in the present
representation implies
\[
\rho^{12} = (\rho^{21})^+
\]
and it relates the upper-diagonal and lower-diagonal components.
This amounts to a single four-dimensional Dirac spinor $\rho^{12}$ and
the model is non-chiral. However, since the fermions transform in complex, bifundamental representations of the gauge group the model does not have a vectorlike structure. The resulting structure corresponds to that of mirror fermions~\cite{Maalampi:1988va}. Let us recall that we have two fuzzy spheres with f\/luxes and we have
assumed already that $m_L>0$ and $m_R>0$.
Then, the relations~(\ref{zeromodes-full-12}) and~(\ref{zeromodes-full-21}) can be
written as
\[
{\chi_{L,{\rm tang}}}|_{\Psi^{12}} = \chi_{R,{\rm tang}}|_{\Psi^{12}} = +1 ,  \qquad
{\chi_{L,{\rm tang}}}|_{\Psi^{21}} = \chi_{R,{\rm tang}}|_{\Psi^{21}} = -1.
\]
It follows from (\ref{zeromodes-full-12}) and (\ref{zeromodes-full-21}) that the operators $\Pi_L$ and $\Pi_R$, def\/ined in (\ref{chiral-map}), actually coincide on the space of zero modes.
Hence the full fermionic Hilbert space can be separated in two sectors as follows,
\begin{gather}
\cH_+=\{\Psi; \, \Pi_L\Psi = \Psi\} \qquad \mbox{and}\qquad
\cH_-=\{\Psi; \, \Pi_L\Psi = -\Psi\}. \label{chiralsectors}
\end{gather}
Then it is clear that $\Psi^{12}$ and $\Psi^{21}$ have opposite four-dimensional
chirality in each sector.
Therefore we end up with two exactly chiral mirror sectors, which are separated according to~(\ref{chiralsectors}).

Therefore our result is that even though the f\/luxes on~$S^2\times S^2$ lead indeed
to the expected zero modes, the model nevertheless
turns out to be non-chiral a priori.
More precisely, we f\/ind essentially mirror models, where
two chiral sectors arise with opposite chirality.
This means that each would-be zero mode from~$\Psi^{12}$
has a mirror partner from~$\Psi^{21}$, with opposite
chirality and gauge quantum numbers.
The reason for this is that the fuzzy geometry
is four-dimensional but in some sense embedded in six extra dimensions.
The missing two (``shadow'') dimensions are ref\/lected
in extra components of the spinors, which do not see the
f\/lux and respectively the chirality on
$S^2\times S^2$. This is a crucial dif\/ference of our model comparing
with models based on commutative extra dimensions,
where chiral Lagrangians are easier to obtain (see
\cite{Kapetanakis:1992hf,Manton:1981es} and the discussion in Section~\ref{section2.4} of the present paper).
Thus we arrive essentially at a picture of mirror fermions
discussed e.g.\ in~\cite{Maalampi:1988va} from a
phenomenological point of  view.
While this may still be physically interesting since the mirror fermions
may have larger mass than the ones we see at low energies,
it would be desirable to f\/ind a chiral version
with similar features. We shall show in the following section that using orbifold techniques this can indeed be achieved.

\section{Orbifolds, fuzzy extra dimensions and chiral models}\label{section4}

In the previous section we discussed that a toroidal dimensional reduction of a ten-dimensional ${\cal N}=1$ SYM theory
to four dimensions leads to ${\cal N}=4$ supersymmetry in four
dimensions, which is not phenomenologically acceptable mainly because it is impossible to accommodate chiral fermions in the theory. The obvious
way to obtain ${\cal N}=1$ four-dimensional models, which might be
realistic since they admit chiral fermions, is to reduce the theory on suitable manifolds such as
Calabi--Yau manifolds~\cite{Candelas:1985en} or manifolds with an
$SU(3)$-structure (see,
e.g.~\cite{LopesCardoso:2002hd,Gauntlett:2003cy}). However, another remarkable way to achieve ${\cal N}=1$ supersymmetry in four dimensions is to perform a reduction on an orbifold \cite{Dixon:1985jw,Bailin:1999nk}.

In this section, in order to pursue further the possibility to obtain chiral low-energy theories within the framework of gauge theories with fuzzy extra dimensions, we shall introduce an orbifold structure similar to the one used in~\cite{Kachru:1998ys}. The authors of~\cite{Kachru:1998ys}, motivated by the celebrated duality between four-dimensional
${\cal N} = 4$, $U(N)$ SYM theory and Type IIB string theory on $AdS_5\times S^5$ \cite{Maldacena:1997re}, used orbifold techniques similar to \cite{Douglas:1996sw,Douglas:1997de} to break some of the four supersymmetries. Considering dif\/ferent embeddings of a $\Z_3$ discrete group in the $R$-symmetry group of the ${\cal N}=4$ SYM theory and performing an orbifold projection of the original theory they determined ${\cal N}=0,1,2$ theories, i.e.\ with reduced supersymmetry. Moreover, the initial gauge group $SU(3N)$ (realised on $3N$ $D3$ branes) is broken down to $SU(N)^3$ and the fermions are accommodated in chiral representations of the gauge group.

\subsection[${\cal N}=4$ SYM and $\Z_3$ orbifolds]{$\boldsymbol{{\cal N}=4}$ SYM and $\boldsymbol{\Z_3}$ orbifolds}\label{section4.1}

In this section we review the basics of the $\Z_3$ orbifold projection of the ${\cal N}=4$ Supersymmetric Yang--Mills (SYM) theory~\cite{Brink:1976bc}. In particular we discuss the action of the discrete group on the various f\/ields of the theory and the resulting superpotential of the projected theory.

Before introducing the orbifold projection, the theory under consideration is the ${\cal N}=4$ supersymmetric $SU(3N)$ gauge theory\footnote{The gauge group is taken to be $SU(3N)$ for notational convenience as it will be clear in the following.}. This theory contains, in ${\cal N}=1$ language, a $SU(3N)$ vector supermultiplet and three adjoint chiral supermultiplets $\F^i$, $i=1,2,3$. The component f\/ields are the $SU(3N)$ gauge bosons $A_{\mu}$, $\mu=1,\dots ,4$, six adjoint real scalars\footnote{In the following we shall often work with the three complex scalars $\f^i$, $i=1,2,3$, which correspond to the complexif\/ication of the six real ones.} $\f^a$, $a=1,\dots, 6$, transforming as $\textbf{6}$ under the $SU(4)_R$ $R$-symmetry of the theory and four adjoint Weyl fermions~$\psi^p$, $p=1,\dots ,4,$ transforming as $\textbf{4}$ under the $SU(4)_R$. The theory is def\/ined on the Minkowski spacetime, whose coordinates are denoted as $x^{\mu}$, $\mu=1,\dots,4$.

 In order to discuss orbifolds we have to consider the discrete group $\Z_3$ generically as a subgroup of $SU(4)_R$. There are three possibilities here, which have a direct impact on the amount of remnant supersymmetry~\cite{Kachru:1998ys}:
\begin{enumerate} \itemsep=0pt
\item $\Z_3$ is maximally in $SU(4)_R$, in which case we are generically led to non-supersymmetric models;
\item $\Z_3$ is embedded in an $SU(3)$ subgroup of the full $R$-symmetry group, leading to $\cN=1$ supersymmetric models with $R$-symmetry $U(1)_R$;
\item $\Z_3$ is embedded in a specif\/ic $SU(2)$ subgroup of $SU(4)_R$, in which case the remaining supersymmetry is ${\cal N}=2$ with $R$-symmetry $SU(2)_R$.
    \end{enumerate}

Let us next discuss in more detail the case where $\Z_3$ is embedded in $SU(3)$, that leads to $\cN=1$ supersymmetric models. In order to proceed we consider a generator $g \in \Z_3$. This generator is conveniently labeled (see \cite{Douglas:1997de}) by three integers $\overrightarrow{a}\equiv (a_1,a_2,a_3)$ which satisfy the condition $a_1+a_2+a_3\equiv 0 \mod 3$. This condition is equivalent to the statement that the discrete group is indeed embedded in $SU(3)$ and therefore it ref\/lects the fact that $\cN=1$ supersymmetry is preserved.

The $\Z_3$ acts non-trivially on the various f\/ields of the theory depending on their transformation properties under the $R$-symmetry.
The geometric action of the $\Z_3$ rotation on the gauge and the gaugino f\/ields is trivial, since they are singlets under $SU(4)_R$.
 On the other hand, the action of $\Z_3$ on the complex scalars is specif\/ied by the matrix  $\g(g)_{ij}=\d_{ij}\o^{a_i}$,  where $\o=e^{\frac{2\pi i}{3}}$, while the corresponding action on the fermions $\psi^i$ is given by  $ \g(g)_{ij}=\d_{ij}\o^{b_i},
$ where\footnote{This relation is of course also understood modulo 3.} $b_i=-\frac 12 (a_{i+1}+a_{i+2}-a_i)$.
In the case under study the three integers have the values $\vec{a}=(1,1,-2)$, which implies
$b_i = a_i$.

However, since the matter f\/ields also transform non-trivially under the gauge group, the discrete group acts on their gauge indices too.
 The action of this rotation can be described by the matrix
\begin{gather}
\label{g3}
\g_3 = \left(\begin{array}{ccc} \one_{N} & 0 & 0 \\
        0 & \omega\one_{N} & 0 \\
        0 & 0 & \omega^2\one_{N} \end{array}\right).
\end{gather}
Let us note that in general the blocks of this matrix could have dif\/ferent dimensionality
(see, e.g.~\cite{Aldazabal:2000sa,Lawrence:1998ja,Kiritsis:2003mc}),
However, anomaly freedom of the projected theory typically
requires that the dimension of the three blocks is the same
as will become obvious in the following.
There is an interesting exception to this rule which will be discussed
in Section~\ref{sec:4222-model}.

In order to derive the projected theory under the orbifold action, one has to keep the f\/ields which are invariant under the combined action of the discrete group on the geometry and on the gauge indices~\cite{Douglas:1997de}. For the gauge bosons the relevant projection is
\[ A_{\mu}=\g_3A_{\mu}\g_3^{-1}.
\]
Therefore, in view of~(\ref{g3}), the gauge group $SU(3N)$ of the original theory is broken down to $H=SU(N)\times SU(N)\times SU(N)$ in the projected theory.

For the complex scalars, which transform non-trivially both under the gauge group and the $R$-symmetry, the projection is
\[
 \f^i_{IJ}=\o^{I-J+a_i}\f^i_{IJ},
  \]
  where $I$, $J$ are gauge indices. This means that $J=I+a_i$ and therefore it is easy to see that the f\/ields which survive the orbifold projection have the form $\f_{I,I+a_i}$ and they transform under the gauge group $H$ as
\begin{gather}
\label{repsH} 3\cdot \bigl((N,\overline{N},1)+(\overline{N},1,N)+(1,N,\overline{N})\bigl).
\end{gather}

For the fermions the situation is practically the same.
More specif\/ically in this case the relevant projection is
\[
 \psi^i_{IJ}=\o^{I-J+b_i}\psi^i_{IJ}.
\]
Then the surviving
fermions have the form $\psi^i_{I,I+b_i}$ and they transform under
$H$ in the representations (\ref{repsH}), exactly as the scalars.
This is just another manifestation of the $\cN=1$ remnant supersymmetry.
Moreover, the structure of the representations (\ref{repsH})
guarantees that the resulting theory does not suf\/fer from any
gauge anomalies\footnote{On the contrary, had we considered that
the matrix~(\ref{g3}) contained blocks of dif\/ferent dimensionality
the projected theory would be anomalous and therefore additional
sectors would be necessary in order to cancel the gauge anomalies.}.

Let us next note two important features of the projected theory. First the fermions transform in chiral representations of the gauge group. Indeed, the representations~(\ref{repsH}) are complex bifundamental ones, and their complex conjugates do not appear in the projected theory. Secondly, there are three fermionic generations in the theory. This is expected since as we noted before the theory contains three chiral supermultiplets under ${\cal N}=1$, leading to three generations.

Concerning the interactions among the f\/ields of the projected theory, let us consider the superpotential of the $\cN=4$ supersymmetric Yang--Mills theory, which has the form~\cite{Brink:1976bc}:
\[
W_{\cN=4} =\epsilon_{ijk}\,{\rm Tr}\,(\F^i\F^j\F^k)
,
\] where the three chiral superf\/ields of the theory appear.
Clearly, the superpotential after the orbifold projection has the same form but it encodes only the interactions among the surviving f\/ields of the resulting ${\cal N}=1$ theory. Therefore it can be written as
\begin{gather}
\label{spot3} W_{{\cal N}=1}^{\rm (proj)}=\sum_{I}\e_{ijk}\F^i_{I,I+a_i}\F^j_{I+a_i,I+a_i+a_j}\F^k_{I+a_i+a_j,I}
,
\end{gather}
where the relation $a_1+a_2+a_3\equiv 0 \mod 3$ was taken into account.

\subsection{Twisted fuzzy spheres}\label{section4.2}

In the present section we introduce the ``twisted fuzzy sphere'' $\tilde S^2_N$, which is
a variant of the ordinary fuzzy sphere \cite{Madore:1991bw} compatible with the orbifolding.
It is def\/ined by the following relations
\begin{gather}
\label{twisted-vacuum}
[\f^i,\f^j]  = i\e_{ijk}(\f^k)^{\dagger}, \qquad \f^i (\f^i)^{\dagger}  =   R^2,
\end{gather}
where $(\f^i)^{\dagger}$ denotes hermitean conjugation of the complex scalar f\/ield $\f^i$
and $[ R^2,\f^i] = 0$.
The relation (\ref{twisted-vacuum}) is compatible with the $\Z_3$ group action, in contrast to the usual fuzzy sphere. Indeed, a quick look at equation~(\ref{repsH}) reveals that the scalar f\/ields are expected to satisfy (\ref{twisted-vacuum}) instead of the commutation relations for the ordinary fuzzy sphere.
Nevertheless the above relations are closely related to a fuzzy sphere.
This can be seen by considering the untwisted f\/ields~$\tilde \f_i$, def\/ined by
\begin{gather}
\f^i = \Omega \tilde\f^i ,
\label{twisted-fields}
\end{gather}
for some $\Omega\ne 1$ which satisf\/ies
\begin{gather}
\Omega^3 = 1, \qquad [\Omega,\f^i] = 0, \qquad \Omega^\dagger = \Omega^{-1}
\label{cond-1}
\end{gather}
and\footnote{Here $[\Omega,\f^i]$ is understood before the orbifolding.}
\begin{gather}
(\tilde\f^i)^\dagger = \tilde\f^i,  \qquad \mbox{i.e.}\qquad (\f^i)^\dagger = \Omega \f^i.
\label{cond-2}
\end{gather}
Then \eqref{twisted-vacuum} reduces to the ordinary fuzzy sphere relation
\begin{gather}
 [\tilde\f^i,\tilde\f^j] = i\e_{ijk}\tilde\f^k,
\label{fuzzy-transf}
\end{gather}
generated by $\tilde\f^i$, as well as to the relation
$
\tilde \f^i\tilde \f^i = R^2.
$
This justif\/ies to call the noncommutative space generated by $\f^i$ a twisted fuzzy sphere.
It is remarkable
that this construction is possible only for $\Z_3$ and for no other
$\Z_n$, thus providing a justif\/ication for our choice of orbifold group.

An interesting realization of a twisted fuzzy sphere (\ref{twisted-vacuum}) is given by
\begin{gather}
\label{solution1}
\phi^i = \Omega\, (\one_3\otimes\lambda^i_{(N)}),
\end{gather} where $\l^i_{(N)}$ denote the generators of $SU(2)$ in the $N$-dimensional irreducible representation and the matrix $\Omega$ is given by
\begin{gather}
\Omega = \Omega_3 \otimes \one_N, \qquad
\Omega_3 = \begin{pmatrix}
		0 & 1 & 0 \\
		0 & 0 & 1 \\
		1 & 0 & 0 \\
\end{pmatrix}, \qquad \Omega^3 = \one.
\label{omega-large}
\end{gather}
The transformation
$\f^i = \Omega \tilde\f^i$
\eqref{twisted-fields} relates the
``of\/f-diagonal'' orbifold sectors~\eqref{repsH} to block-diagonal conf\/igurations as follows,
\begin{gather}
\label{transuntwist}
\f^i = \begin{pmatrix}
		0 & (\l^i_{(N)})_{(N,\overline N,1)} & 0 \\
		0 & 0 & (\l^i_{(N)})_{(1,N,\overline N)} \\
		(\l^i_{(N)})_{(\overline N,1,N)} & 0 & 0 \\
\end{pmatrix}
  =   \Omega \begin{pmatrix}
		\l^i_{(N)} & 0 & 0\\
		 0 & \l^i_{(N)} & 0 \\
		0 & 0 & \l^i_{(N)}  \\
\end{pmatrix}.
\end{gather} We observe that the untwisted f\/ields~$\tilde \f^i$, which generate the fuzzy sphere, acquire a block-diagonal form. Each one of these blocks satisf\/ies separately the fuzzy sphere relation~(\ref{fuzzy-transf}) and therefore it is natural to reinterpret this conf\/iguration as three fuzzy spheres of fuzziness~$N$.
The solution~$\f^i$ can thus
be interpreted as twisted conf\/iguration of three fuzzy spheres
compatible with the orbifolding.

The solution (\ref{solution1}) breaks completely the gauge symmetry~$SU(N)^3$.
This geometrical interpretation
is helpful to understand the f\/luctuations
around these fuzzy orbifolds. However, for our purposes it will be useful to consider solutions which do not break the~$SU(N)^3$ gauge symmetry completely but they break it down to a smaller gauge group. We shall study such solutions in the following paragraph and present specif\/ic applications in the upcoming sections.

\subsection{Dynamical generation of twisted fuzzy spheres}
\label{section4.3}

Let us now show how the above geometries can arise as a vacuum solution of the
f\/ield theory which was considered in Section~\ref{section4.1}. As it was previously described, the superpotential of the theory after the orbifold projection has the form~(\ref{spot3}). Therefore one can easily read of\/f the corresponding potential, which is\footnote{Here we restrict to the scalar sector, since this is the relevant one for the search of fuzzy sphere vacua. Moreover, the gauge indices are suppressed.}
\[
V_{{\cal N}=1}^{(proj)}(\f)= \frac 14 \,{\rm Tr}\,\big([\f^i,\f^j]^\dagger [\f^i,\f^j]\big),
\]
where $\f^i$ denotes the scalar component of the superf\/ield $\F^i$. The minimum of this potential is obtained for vanishing vevs of the f\/ields and therefore vacua corresponding to non-commutative geometries do not exist without any additional modif\/ications.

Clearly, in order to determine a minimum of the potential of the form (\ref{twisted-vacuum}) we have to make the following modif\/ications in the theory. First of all, we have to add ${\cal N}=1$ soft supersymmetry breaking
(SSB) terms of the form\footnote{Here we present a set of scalar SSB terms. However, there exist of
course other soft terms such as $\frac 12 M\l\l$, where $\l$ is the gaugino and $M$ its mass,
which has to be included in the full SSB sector~\cite{Djouadi:2005gj}.}
\begin{gather}\label{soft}
V_{\rm SSB}=\frac 12 \sum_i m^2_i  {\f^i}^{\dagger}\f^i+\frac 12 \sum_{i,j,k} h_{ijk}\f^i\f^j\f^k+{\rm h.c.},
\end{gather}
where $h_{ijk}$ vanishes unless $i+j+k\equiv 0\mod 3$.
Of course
a set of SSB terms in the potential is necessary anyway in order for the theory to have a
chance to become realistic, see e.g.~\cite{Djouadi:2005gj}. After the addition of these soft terms as well as of the $D$-terms the
full potential of the theory becomes
\begin{gather}\label{potential1}
V=V_{{\cal N}=1}^{\rm (proj)}+V_{\rm SSB}+V_{D},
\end{gather}
where $V_{D}=\frac 12 D^2=\frac 12 D^{I}D_{I}$ includes the $D$-terms of the theory.
These $D$-terms have the form $D^{I}=\f_i^{\dagger}T^{I}\f^i$, where $T^I$ are the generators
of the representation of the corresponding chiral multiplets.

In order to allow for twisted
fuzzy sphere vacua, we now make the choice $h_{ijk}=\epsilon_{ijk}$ and $m_i^2=1$.
A more general possibility will be investigated in Section~\ref{section4.5}. Then the potential~(\ref{potential1}) can be brought in the form
\[
V=\frac 14 (F^{ij})^{\dag}F^{ij}   + V_D,
\]
where we have def\/ined
\begin{gather}
\label{fieldstrength}
F^{ij}=[\f^i,\f^j]-i\e^{ijk}(\f^k)^{\dagger}.
\end{gather}
The f\/irst term of the potential is positive def\/inite, and vanishes if the
relation~(\ref{twisted-vacuum}) holds. Therefore the global minimum of the potential
is realized by a twisted fuzzy sphere $\tilde S^2_N$ \eqref{twisted-vacuum},
at least for a suitable range of parameters
in the potential. The quartic term~$V_D$ will typically only modify
its radius, as in the case of the ordinary fuzzy sphere~\cite{Aschieri:2006uw,Steinacker:2003sd}.
 The expression~(\ref{fieldstrength}) will be interpreted in the
following as the f\/ield strength on the spontaneously generated fuzzy extra dimensions. Let us note that in general the potential
may have several dif\/ferent local minima,
which may be given e.g.\ by twisted fuzzy spheres with various radii; we will not discuss
possible meta-stable vacua or phase-transitions here.

Let us now study further the vacuum and its geometric interpretation.
The scalar f\/ields $\phi^i$ are governed by the potential~(\ref{potential1}),
which includes the F- and D-terms as well as the SSB terms.
Under suitable conditions,
this potential clearly has a twisted fuzzy sphere
solution
\begin{gather}
\phi^i = \Omega \bigl(\one_3\otimes(\lambda^i_{(N-n)} \oplus 0_{n})\bigl),
\label{twistedfuzzys2-2}
\end{gather}
where $0_{n}$ denotes the $n\times n$ matrix with vanishing entries.
The gauge symmetry is broken from $SU(N)^3$ down to $SU(n)^3$.
This vacuum should be interpreted as $\R^4 \times \tilde S^2_N$
with a twisted
fuzzy sphere in the $\phi_i$ coordinates.

In order to understand the f\/luctuations of the scalar f\/ields around
this vacuum, the transformation $\phi^i = \Omega\tilde\phi^i$ is useful.
Fluctuations around the ordinary fuzzy sphere
$S^2_N$ are known to describe gauge and scalar
f\/ields on $S^2_N$ \cite{Madore:2000en,Steinacker:2003sd},
and in particular they all become massive
from the point of view of $\R^4$.
We have seen in
(\ref{transuntwist}) that the twisted sphere $\tilde S^2_N$ is mapped by $\Omega$ into three fuzzy spheres $\tilde \f^i$
embedded in the diagonal
$N\times N$ blocks of the original $3N\times 3N$ matrix.
Therefore all f\/luctuations can be understood as f\/ields on the
three diagonally embedded untwisted fuzzy spheres:
\[
\tilde \phi^i = \lambda^i_{(N)} + A^i ,
\]
and the f\/ield strength \eqref{fieldstrength} reduces to the f\/ield strength on a
fuzzy sphere
\[
F^{ij}=[\f^i,\f^j]-i\e^{ijk}(\f^k)^{\dagger}
= \Omega^2 ([\tilde\f^i,\tilde\f^j]-i\e^{ijk}\tilde\f^k)
\]
as long as \eqref{cond-1} and \eqref{cond-2} hold.
The vacuum can thus be interpreted at intermediate energy scales
as $\R^4 \times S^2_N$
with three (untwisted) fuzzy spheres
in the $\tilde\phi_i$  coordinates.
Moreover, due to the orbifolding condition
there are no of\/f-diagonal components relating these dif\/ferent
spheres. It now follows as in \cite{Aschieri:2006uw,Steinacker:2007ay}
that the gauge f\/ields and fermions can be
decomposed into Kaluza--Klein towers of massive modes
on $S^2_N$ resp.~$\tilde S^2_N$
due to the Higgs ef\/fect,
as well as a massless sector.

\subsection{Chiral models from the fuzzy orbifold}
\label{section4.4}

In this section we discuss three particular models which can be constructed in the above context.
In all cases we start by considering the ${\cal N}=4$ SYM theory in four dimensions with gauge group $SU(3N)$. As we have already mentioned this theory contains, in ${\cal N}=1$ language, an $SU(3N)$ vector supermultiplet and three adjoint chiral supermultiplets $\Phi^i$ with superpotential
\[
 W_{{\cal N}=4}=\epsilon_{ijk}\,{\rm Tr}\,(\F^i\F^j\F^k).
\]

Subsequently we choose the discrete group $\Z_3$ and embed it in the $SU(3)$ part of the $R$-symmetry. Performing the orbifold projection, as it was described in Section~\ref{section2}, we obtain an ${\cal N}=1$ theory with vectors in $SU(N)^3$ and complex scalars and fermions in chiral representations of the gauge group. In particular, according to (\ref{repsH}), there are three families, each transforming under the gauge group $H$ as
\begin{gather}
\label{repsHmodel}
(N,\overline{N},1)+(\overline{N},1,N)+(1,N,\overline{N}).
\end{gather}
Moreover, the superpotential takes the form~(\ref{spot3}). The dif\/ference between the models lies in the next step of the construction, where the gauge group $SU(N)^3$ will be broken spontaneously to a unif\/ication group. The minimal cases which satisfy the requirement of anomaly freedom are the gauge groups $SU(4)\times SU(2)\times SU(2), SU(4)^3$ and $SU(3)^3$.

\subsubsection[A $SU(4)_c\times SU(2)_L\times SU(2)_R$ model]{A $\boldsymbol{SU(4)_c\times SU(2)_L\times SU(2)_R}$ model}
\label{sec:4222-model}

In order to obtain the Pati--Salam gauge group $SU(4)_c\times SU(2)_L\times SU(2)_R$~\cite{Pati:1974yy} (see also~\cite{Antoniadis:1988cm} and~\cite{Anastasopoulos:2010ca} for a recent study), we decompose the integer $N$ in two dif\/ferent ways, namely as
\[
 N = n_1+4, \qquad  N = n_2+2.
 \]
  Then we consider the following regular embeddings,
\[
SU(N) \supset SU(n_1)\times SU(4)\times U(1), \qquad
SU(N) \supset SU(n_2)\times SU(2)\times U(1).
\]
The full gauge group is accordingly decomposed as
\[
 SU(N)^3\supset SU(n_1)\times SU(4)\times SU(n_2)\times SU(2)\times SU(n_2)\times SU(2)\times U(1)^3.
 \]
Performing a shuf\/f\/ling of the group factors and ignoring the $U(1)$s\footnote{These may be anomalous and become massive by the Green--Schwarz mechanism and therefore they decouple at low energies~\cite{Lawrence:1998ja}.} it is easy to see that the original representations~(\ref{repsH}) are decomposed as follows,
\begin{gather*}
  SU(n_1)\times SU(n_2)\times SU(n_2)\times SU(4)\times SU(2)\times SU(2),  \\
  (n_1,\obar{n}_2,1;1,1,1)+(1,n_2,\obar{n}_2;1,1,1)+(\obar{n}_1,1,n_2;1,1,1)+
 (1,1,1;4,2,1)+(1,1,1;1,2,2)\\
 \qquad {} +(1,1,1;\obar{4},1,2)
 +(n_1,1,1;1,2,1)+(1,n_2,1;1,1,2)+(1,1,n_2;\obar{4},1,1) \\
 \qquad{}
  +(\obar{n}_1,1,1;1,1,2)+(1,\obar{n}_2,1;4,1,1)+(1,1,\obar{n}_2;1,2,1).
  \end{gather*}
First of all it is important to note that the theory is anomaly free. This is merely due to the special feature of $SU(2)$, where the fundamental representation is self-conjugate. Therefore, although the structure involves a product of dif\/ferent gauge groups, it is still not anomalous.

Now utilizing the mechanism of Section~\ref{section3}, fuzzy extra dimensions can be dynamically generated and the unbroken gauge group at low-energies is $SU(4)_c\times SU(2)_L\times SU(2)_R$, with f\/ields transforming under the representations
\begin{gather*}
 SU(4)\times SU(2)\times SU(2),\\
		  3\cdot \bigl((4,2,1)+(\overline{4},1,2)+(1,2,2)\bigl).
\end{gather*}
This is realized by the following vacuum
\begin{gather*}
\phi^i = \Omega \bigl(0_2 \oplus \one_3\otimes (\lambda^i_{(N-2)} \oplus 0_{2})\bigl),
\qquad
\Omega = \begin{pmatrix}
		\one_2 & 0  \\
		0 & \Omega_3\otimes\one_{N}
\end{pmatrix}
\end{gather*}
interpreted in terms of twisted fuzzy spheres,
where $\Omega_3$ is def\/ined in \eqref{omega-large}.

Then the quarks and leptons of the SM f\/it in these representations. For example, the f\/irst generation is represented as
\[ 
f\sim(4,2,1)  =   \left(\begin{array}{cc} d_L^1 & u_L^1 \\ d_L^2 & u_L^2 \\ d^3_L & u^3_L \\ e_L & \nu_L
\end{array}\right) ,  \qquad
f^c\sim(\bar 4,1,2)  =   \left(\begin{array}{cccc} d^{1 c}_L & d^{2 c}_L &
d^{3 c}_L & e_L^c
\\
u^{1 c}_L & u^{2 c}_L & u^{3 c}_L & \nu^c_L \end{array}\right),
\]
 and similarly for the other two generations. Moreover, the $h\sim(1,2,2)$ representation involves the Higgses and the Higgsini.

\subsubsection[A $SU(4)_c\times SU(4)_L\times SU(4)_R$ model]{A $\boldsymbol{SU(4)_c\times SU(4)_L\times SU(4)_R}$ model}

A further possibility is the gauge group $SU(4)_c\times SU(4)_L\times
SU(4)_R$, where $SU(4)_c$ is again the Pati--Salam colour gauge group.
This gauge group can be obtained by decomposing $N$ as
\[ N=n+4,
\]
leading to the decomposition of $SU(N)^3$ to $SU(n)^3\times SU(4)^3$ with particle content
\begin{gather*}
 SU(n)\times SU(n)\times SU(n)\times SU(4)\times SU(4)\times SU(4),\\
  (n,\obar{n},1;1,1,1)+(1,n,\obar{n};1,1,1)+(\obar{n},1,n;1,1,1)
  +(1,1,1;4,\obar{4},1)\\
  \qquad{}
  +(1,1,1;1,4,\obar{4})+(1,1,1;\obar{4},1,4)
  +(n,1,1;1,\obar{4},1)+(1,n,1;1,1,\obar{4})\\
  \qquad{} +(1,1,n;\obar{4},1,1)
  +(\obar{n},1,1;1,1,4)+(1,\obar{n},1;4,1,1)+(1,1,\obar{n};1,4,1).
 \end{gather*}

This is realized by the following vacuum, interpreted
in terms of twisted fuzzy spheres $\tilde S^2_{N-4}$
as in \eqref{twistedfuzzys2-2}:
\begin{gather*}
\phi^i = \Omega\bigl(\one_3\otimes (\lambda^i_{(N-4)} \oplus 0_{4})\bigl),
\end{gather*}
where $\Omega$ is def\/ined in \eqref{omega-large}.
Decomposing $SU(N) \supset SU(n)\times SU(4)\times U(1)$,
the gauge group is broken to $SU(4)^3$, and the low-energy f\/ield
content
is
\begin{gather}
  SU(4)\times SU(4)\times SU(4),\nonumber\\
		  3\cdot \bigl((4,\overline{4},1)+(\overline{4},1,4)+(1,4,\overline{4})\bigl).\label{repsfinal3}
\end{gather}
This case has been examined originally in~\cite{Ibanez:1998xn} and from a phenomenological viewpoint in \cite{Ma:2004mi}. The quarks and leptons of the f\/irst family should transform as
\begin{gather*}
f = \left(\begin{array}{cccc} d & u & y & x \cr d & u & y & x \cr d & u & y & x \cr
e & \nu & a & v \end{array}\right) \sim (4,\overline 4,1), \qquad f^c = \left(\begin{array}{cccc} d^c & d^c & d^c & e^c
\cr u^c & u^c & u^c & \nu^c \cr y^c & y^c & y^c & a^c \cr x^c & x^c & x^c &
v^c \end{array}\right)\sim (\overline 4,1,4).
\end{gather*}
Clearly, there have to be new heavy quarks and leptons and in addition the supermultiplet $h\sim(1,4,\overline{4})$ still has to be considered.

A very interesting feature which we would like to point out here is that the one-loop $\beta$-function coef\/f\/icient in the renormalization group equation of each $SU(4)$ gauge coupling is given~by
\begin{gather*}
 b=\left(-\frac{11}{3}+\frac 23\right)\cdot3+n_{f}\left(\frac 23+\frac 13\right)\cdot\frac 12 \cdot2\cdot 3,
  \end{gather*}
  which for the present case of $n_f=3$ copies of the supermultiplet (\ref{repsfinal3}) results in
  \[
   b=0.
   \]
Therefore, we observe that the existence of three families of quarks and leptons leads to one of the necessary conditions for a f\/inite f\/ield theory. Let us mention that this is a general feature of models with a $SU(N)^k$ gauge group, independently of the values of~$N$ and~$k$~\cite{Ma:2004mi}. Therefore it also holds in the following case of $SU(3)^3$.

\subsubsection[A $SU(3)_c\times SU(3)_L\times SU(3)_R$ model]{A $\boldsymbol{SU(3)_c\times SU(3)_L\times SU(3)_R}$ model}\label{section4.4.3}

Let us now turn to another possibility, the trinif\/ication group $SU(3)_c\times SU(3)_L\times SU(3)_R$ \cite{Glashow:1984gc,Rizov:1981dp}, which was also studied in \cite{Ma:2004mi,Lazarides:1993uw,Heinemeyer:2009zs,Babu:1985gi,Leontaris:2005ax} and from a string theory perspective in \cite{Kim:2003cha}. In the present case we consider the following picture. Let us decompose the integer $N$ as
\[
 N=n+3.
 \] Subsequently, let us consider the regular embedding
\begin{gather}
\label{dec3}
SU(N)\supset SU(n)\times SU(3)\times U(1).
\end{gather}
  Then the relevant embedding for the full gauge group is
\[
 SU(N)^3\supset SU(n)\times SU(3)\times SU(n)\times SU(3)\times SU(n)\times SU(3)\times U(1)^3.
 \]
 The three $U(1)$ factors decouple from the low-energy sector of the theory, as it was mentioned above. The representations~(\ref{repsHmodel}) are then decomposed accordingly (notice the shuf\/f\/ling in the group factors),
\begin{gather}
 SU(n)\times SU(n)\times SU(n)\times SU(3)\times SU(3)\times SU(3),\nonumber\\
  (n,\obar{n},1;1,1,1)+(1,n,\obar{n};1,1,1)+(\obar{n},1,n;1,1,1)
  +(1,1,1;3,\obar{3},1)\nonumber\\
  \qquad{} +(1,1,1;1,3,\obar{3})+(1,1,1;\obar{3},1,3)
  +(n,1,1;1,\obar{3},1)+(1,n,1;1,1,\obar{3})\nonumber\\
  \qquad{} +(1,1,n;\obar{3},1,1)
  +(\obar{n},1,1;1,1,3)+(1,\obar{n},1;3,1,1)+(1,1,\obar{n};1,3,1).\label{repsall}
\end{gather}
This is realized by the following vacuum, interpreted
in terms of twisted fuzzy spheres $\tilde S^2_{N-3}$
as in \eqref{twistedfuzzys2-2}:
\begin{gather}
\phi^i = \Omega \big[\one_3\otimes \big(\lambda^i_{(N-3)} \oplus 0_{3}\big)\big].
\label{twistedfuzzys2-4}
\end{gather}
Considering the decomposition (\ref{dec3}),
the gauge group is broken
to $K = SU(3)^3$.
Finally, the surviving f\/ields under the unbroken gauge group $K$ transform in the following representations,
\begin{gather}
   SU(3)\times SU(3)\times SU(3),\nonumber\\
		 3\cdot \bigl((3,\overline{3},1)+(\overline{3},1,3)+(1,3,\overline{3})\bigl).\label{repsfinal}
\end{gather}
These are the desired chiral representations of the unif\/ication group $SU(3)_c\times SU(3)_L\times SU(3)_R$.
The quarks of the f\/irst family transform under the gauge group as
\begin{gather}\label{quarks3}
q = \left(\begin{array}{ccc} d & u & h \cr d & u & h \cr d & u & h \end{array}\right)\sim (3,\overline 3,1), \qquad
q^c =\left(\begin{array}{ccc} d^c & d^c & d^c \cr u^c & u^c & u^c \cr h^c & h^c & h^c \end{array}\right)
\sim (\overline 3,1,3),
\end{gather}
and the leptons transform as
\begin{gather}\label{leptons3}
\lambda = \left(\begin{array}{ccc} N & E^c & \nu \cr E & N^c & e \cr \nu^c & e^c & S  \end{array}\right)
\sim (1,3,\overline 3).
\end{gather}
Similarly, the corresponding matrices for the quarks and leptons of the other two families can be written down.

\subsubsection{A closer look at the masses}

A vital issue of our construction is whether there exist massless and massive modes at the same time. Clearly we need both of these sets; the massless modes in order to obtain chiral fermions and the massive modes in order to reproduce the Kaluza--Klein tower and provide undoubtful justif\/ication that the theory develops fuzzy extra dimensions.

A way to see this through the embeddings we presented before is the following. Let us work out the case of $SU(3)^3$, since the same arguments apply to the other two cases as well. Under the f\/inal gauge group $SU(3)^3$ the fermions transform in the representations (\ref{repsfinal}), hence they are chiral. Therefore they remain massless since they are protected by chiral symmetry.

On the other hand, looking at (\ref{repsall}) we can make two crucial observations. First of all, it becomes clear from the vacuum solution (\ref{twistedfuzzys2-4}) that the scalar f\/ields which acquire vevs in this vacuum are the following,
\[ \langle(n,\obar{n},1;1,1,1)\rangle, \qquad \langle(1,n,\obar{n};1,1,1)\rangle, \qquad \langle(\obar{n},1,n;1,1,1)\rangle.
\]
Then all the fermions, apart from the chirally protected ones, obtain masses, since we can form the invariants
\begin{gather} 
(1,n,\obar{n};1,1,1)\langle(n,\obar{n},1;1,1,1)\rangle (\obar{n},1,n;1,1,1)+ \mbox{cyclic permutations},\nonumber\\
\label{inv2}
(\obar{n},1,1;1,1,3)\langle(n,\obar{n},1;1,1,1)\rangle (1,n,1;1,1,\obar{3}) \qquad \mbox{etc.},
\end{gather}
and the corresponding ones for all the other fermions. In these invariants the f\/ield in the middle is the scalar f\/ield which acquires the vev~\eqref{twistedfuzzys2-4},
while the other two are fermions, i.e.\ the invariants are trilinear Yukawa terms and they are responsible for the fermion masses after the spontaneous symmetry breaking. Therefore a f\/inite Kaluza--Klein tower of massive fermionic modes appears,
consistent with the interpretation of the vacuum~(\ref{twistedfuzzys2-4}) as a higher-dimensional theory with spontaneously generated fuzzy extra dimensions. In particular, the f\/luctuations from this vacuum correspond to the internal components of the higher-dimensional gauge f\/ield. Also, as far as the fermions transforming as $(1,1,1;3,\overline 3,1)$, $(1,1,1;\overline 3,1,3)$ and $(1,1,1;1,3,\overline 3)$ are concerned, obviously there does not exist any trilinear invariant that they could form with one of the scalar f\/ields which acquire a vev. Therefore, as it was already mentioned, they remain massless and they are the chiral fermions of the model.

Finally, it is worth noting that in (\ref{inv2}) the ``internal'' structure and the ``observable'', low-energy structure appear mixed and therefore these Kaluza--Klein fermion masses may have an ef\/fect on the $SU(3)^3$
phenomenology \cite{Dienes:1998vh,Ghilencea:1998st,Kobayashi:1998ye,Kubo:1999ua}.

\subsection[Fuzzy breaking for $SU(3)^3$]{Fuzzy breaking for $\boldsymbol{SU(3)^3}$}\label{section4.5}

In this section we discuss another possible application of the fuzzy orbifold construction which was presented above. We focus on the most interesting case\footnote{The breaking of the models with gauge group $SU(4)\times SU(2)\times SU(2)$ and $SU(4)^3$ was studied along the same lines in~\cite{Chatzistavrakidis:2010xi}, where it was shown that they meet serious phenomenological obstacles.} of $SU(3)^3$ and the orbifold projection is utilized to study its spontaneous breaking down to the MSSM and the $SU(3)_c\times U(1)_{\rm em}$. It is important to note that we shall focus only on symmetry breaking patterns where additional superf\/ields are not introduced, namely the model is broken spontaneously due to its own scalar sector.

In particular, instead of starting with a $SU(3N)$ gauge theory with a large $N$, we can start with a smaller gauge group, namely $SU(9)$, in order to obtain the model $SU(3)^3$ after orbifolding. Therefore the initial set-up
consists of the ${\cal N}=4$ SYM theory with gauge f\/ields in the gauge group $SU(9)$. Subsequently a $\Z_3$ orbifold projection is performed in the spirit of Section~\ref{section4.1}.

Alternatively, this procedure may be viewed as a second step of the construction which was presented in Section~\ref{section4.4.3}. Indeed, if such a view is adopted, after the large-$N$ symmetry breaking, the $SU(3)^3$ model is obtained. It involves a superpotential and the corresponding soft supersymmetry breaking terms. Part of the SSB sector is naturally inherited from the corresponding one in the large-$N$ model, namely it is already contained in the expression (\ref{soft}) for suitable $h_{ijk}$. This fact justif\/ies further the use of the same technique, the spontaneous generation of twisted fuzzy spheres, in order to achieve the spontaneous symmetry breaking down to the MSSM and subsequently to the $SU(3)\times U(1)_{\rm em}$.

Specif\/ically, the model is obtained from an $SU(9)$ gauge theory as follows.
We perform an orbifold projection of the ${\cal N}=4$, $SU(9)$ SYM theory such that the $\g_3$ of equation~(\ref{g3}) takes the form
\[ \g_3 = \left(\begin{array}{ccc} \one_{3} & 0 & 0 \\
        0 & \omega\one_{3} & 0 \\
        0 & 0 & \omega^2\one_{3} \end{array}\right).
        \]
Then, according to the rules of Section~\ref{section2}, the gauge group breaks down to $SU(3)^3$, with three chiral supermultiplets transforming as
\begin{gather*}
  SU(3)\times SU(3)\times SU(3),\\
  3\cdot \bigl((3,\overline 3,1)+(\overline{3},1,3)+(1,3,\overline 3)\bigl).
\end{gather*}

First of all, the quarks of the f\/irst family transform under the gauge group as in equation~(\ref{quarks3}) and the leptons transform as in equation~(\ref{leptons3}). The superpotential~(\ref{spot3}) after the orbifold projection in this case becomes~\cite{Ma:2004mi}
\[
W_{{\cal N}=1}^{\rm (proj)}(\l,q,q^c)=YTr(\l q^cq)
+Y'\e_{ijk}\e_{abc}\big(\l_{ia}\l_{jb}\l_{kc}+q^c_{ia}q^c_{jb}q^c_{kc}
+q_{ia}q_{jb}q_{kc}\big),
\]
where the family superscripts are suppressed.
The last terms are special in the $SU(3)^3$ case, and may involve dif\/ferent
families.
The soft supersymmetry breaking terms, which are necessary in order to obtain vacua in the form of twisted fuzzy spheres, are correspondingly read of\/f from equation~(\ref{soft}) with the appropriate $h_{ijk}$ in order to incorporate dif\/ferent scales for the GUT and the EW symmetry breaking\footnote{Of course the EW symmetry breaking of the MSSM requires the introduction of extra soft supersymmetry breaking terms (see e.g.~\cite{Djouadi:2005gj}).}.

The spontaneous breaking of this unif\/ication model down to the MSSM has been studied in several publications \cite{Ma:2004mi,Lazarides:1993uw,Babu:1985gi} and it can be achieved in dif\/ferent ways. Here we would like to mention that in all the known symmetry breaking patterns either additional superf\/ields have to be introduced in the theory \cite{Lazarides:1993uw} or the breaking has to happen in more than one steps, e.g.\ through the left-right symmetric model $SU(3)\times SU(2)_L\times SU(2)_R\times U(1)_{L+R}$ \cite{Ma:2004mi}.

Here we would like to present a dif\/ferent symmetry breaking pattern, where the initial $SU(3)^3$ gauge symmetry is spontaneously broken due to the existing scalar sector of the model, i.e.\ without the need of any additional superf\/ields, and moreover the breaking happens in one step. In order to achieve this we shall utilize the fuzzy orbifold techniques which were presented previously.

Let us recall that the f\/ields of one family can be represented by the following matrix,
\[
\left(\begin{array}{ccc}
0_3 & q & 0_{3}  \\
0_{3} & 0_3 & \l \\
q^c & 0_{3} & 0_3
\end{array}\right),
\] where $0_3$ is the $3\times 3$ matrix with all the entries zero.
Obviously the quark blocks cannot acquire a vev, since this would
break the colour $SU(3)$ gauge group factor.
Therefore the term $Tr (\lambda q q^c)$ in the superpotential
cannot play any role here.
The block which corresponds to the lepton supermultiplet may acquire
vevs only in the directions which have zero hypercharge. This means
that out of the nine components of this block only f\/ive may acquire a
vev, namely $S$, $\nu$, $\nu^c$, $N$ and $N^c$. The f\/irst three are responsible
for the breaking down to the MSSM, while the last two take care of the
EW breaking. Such a vacuum may indeed arise here
due to the presence of the
$\e_{ijk}\e_{abc} \l_{ia}\l_{jb}\l_{kc}$ term in the
superpotential, and moreover we can
interpret it again in terms of a twisted fuzzy sphere.

To see the relation with a twisted fuzzy sphere
(\ref{twisted-vacuum}), we transform the
lepton matrices as $\l'^i = \Omega_3 \lambda^i$ where
$\Omega_3=
\left(\begin{array}{ccc}
0 & 1 & 0  \\
0 & 0 &1 \\
1 & 0 & 0
\end{array}\right)$, noting that the relevant term
$\e_{ijk}\e_{abc} \l_{ia}\l_{jb}\l_{kc}$ is invariant
(up to sign) under such a
transformation. Then $\l$ is transformed to
\[ \l'=
\left(\begin{array}{ccc}
E & N^c & e  \\
\nu^c & e^c & S \\
N & E^c & \nu
\end{array}\right).
\]
Now consider a vacuum solution of the form (superscripts here denote families):
\begin{gather} \label{l1}
\l'^1 = \left(\begin{array}{ccc}
0&k_1 & 0 \\ 0&0&0 \\0&0&0
\end{array}\right), \qquad 
\l'^2 = \left(\begin{array}{ccc}
0 & 0&0\\ 0&0&k_2\\ 0&0&0
\end{array}\right),\qquad 
\l'^3 = \left(\begin{array}{ccc}
0 & 0&0\\ 0&0&0\\ k_3&0&0
\end{array}\right),
\end{gather} while everything else acquires a vanishing vev. These vevs
correspond to the directions of~$N$,~$N^c$ and $S$. The above matrices
satisfy
\[ [\l'^i,\l'^j]=ih_{ijk}(\l'^k)^{\dagger}, \]
where we have def\/ined again
\[ h_{ijk}\equiv \frac{k_ik_j}{k_k}\e_{ijk}. \] This is a
generalization of the twisted fuzzy sphere vacuum where more than one
scales may be included. In the present model this is desirable, since
at least two scales have to be introduced, corresponding to the GUT
and EW breaking respectively.
Moreover, since the model enjoys ${\cal N}=1$ supersymmetry these scales may in principle remain separate.

On the other hand, if we
transform the
lepton matrices as
\[
\l''^i = \Omega'_3 \lambda^i,\qquad \mbox{where}\qquad \Omega'_3=
 \left(\begin{array}{ccc}
0 & 1 & 0  \\
 1 & 0 &0 \\
 0 & 0 & 1
 \end{array}\right),
\qquad \mbox{then}\qquad
\l''=
\left(\begin{array}{ccc}
E & N^c & e  \\
N & E^c & \nu \\
\nu^c & e^c & S
\end{array}\right).
\]
The same twisted fuzzy sphere vacuum as before, namely the matrices (\ref{l1}), 
corresponds now to the directions $\nu$, $\nu^c$ and $N^c$. Therefore, with the above procedure all the neutral directions acquire a vev and the original $SU(3)^3$ model is spontaneously broken down to $SU(3)_c\times U(1)_{\rm em}$. In particular, at the scale where the directions $\nu,\nu^c$ and $S$ acquire vevs, $SU(3)^3$ is spontaneously broken down to the MSSM. Subsequently, at the scale where the $N$ and $N^c$ directions acquire vevs the breaking down to $SU(3)_c\times U(1)_{\rm em}$ takes place. As we have already mentioned these scales are hopefully kept separate by supersymmetry. In other words the hierarchy problem is the same as in any supersymmetric particle physics model.

The remarkable new result of the above procedure is that the spontaneous breaking of the $SU(3)^3$ model acquires an interesting geometrical explanation. It takes place solely due to the Higgsing of the twisted fuzzy spheres in the extra dimensions, without the need of any additional superf\/ields and without the need of any intermediate breaking.

\section{Discussion and conclusions}\label{section5}

Non-commutative geometry has been regarded as a promising framework
for constructing f\/inite quantum f\/ield theories or at least as a natural
scheme for regularizing quantum f\/ield theories. However the quantization
of f\/ield theories on non-commutative spaces has turned out to be much
more dif\/f\/icult than expected and with problematic ultraviolet features~\cite{Filk:dm,Minwalla:1999px}, see however~\cite{Grosse:2005da}, and~\cite{Steinacker:2003sd,Grosse-Steinacker}.

Although SM type of models have been constructed using the Seiberg--Witten
map, they can only be considered as ef\/fective theories and they are not
renormalizable. A drastic change in the perspective of non-commutative
geometry was given with the suggestion that indeed it might be
relevant for particle physics models but in the description of extra
dimensions~\cite{Aschieri:2003vy}. The higher-dimensional theories that can be constructed
based on this proposal, reviewed in the present article, appeared
to have many interesting unexpected features ranging from their
renormalizability to their predictivity.

In the f\/irst part of the review we have considered ideas from non-commutative geometry in order to construct particle physics models which could turn out to be phenomenologically viable. Specif\/ically, in the context of higher-dimensional gauge theories, we explored the possibility that the extra dimensions are described by matrix approximations to smooth manifolds known as fuzzy spaces. Let us now summarize our results and discuss prospects of further work on this subject.

\looseness=1
In the f\/irst part of this paper we considered higher-dimensional gauge theories def\/ined on the product of Minkowski space and a fuzzy coset space $(S/R)_F$ and their dimensional reduction to four dimensions using the CSDR scheme. Although for the technicalities one has to consult the detailed exposition in Section~\ref{section2}, a major dif\/ference between fuzzy and ordinary CSDR is that in the fuzzy case one always embeds $S$ in the gauge group $G$ instead of embedding just $R$ in $G$. This is due to the fact that the
dif\/ferential calculus on the fuzzy coset space is based on $\dim S$
derivations instead of the restricted $\dim S - \dim R$ used in the
ordinary one.  As a result the four-dimensional gauge group $H =
C_G(R)$ appearing in the ordinary CSDR after the geometrical
breaking and before the spontaneous symmetry breaking due to the
four-dimensional Higgs f\/ields does not appear in the fuzzy CSDR. In
fuzzy CSDR the spontaneous symmetry breaking mechanism takes already
place by solving the fuzzy CSDR constraints. The four-dimensional
potential has the typical ``mexican hat'' shape, but it appears
already shifted to a minimum. Therefore in four dimensions appears
only the physical Higgs f\/ield that survives after a spontaneous
symmetry breaking. Correspondingly in the Yukawa sector of the
theory we have the results of the spontaneous symmetry breaking,
i.e.\ massive fermions and Yukawa interactions among fermions and the
physical Higgs f\/ield. Having massive fermions in the f\/inal theory is
a generic feature of CSDR when $S$ is embedded in~$G$~\cite{Kapetanakis:1992hf}. We see that if one would like to describe the
spontaneous symmetry breaking of the SM in the present framework,
then one would be naturally led to large extra \mbox{dimensions}.

A fundamental dif\/ference between the ordinary CSDR and its fuzzy
version is the fact that a non-Abelian gauge group $G$ is not really
required in high dimensions. Indeed  the presence of a $U(1)$ in the
higher-dimensional theory is enough to obtain non-Abelian gauge
theories in four dimensions.


In a further development, we have presented a renormalizable four-dimensional $SU(N)$ gauge theory with a suitable multiplet of
scalars, which dynamically develops fuzzy extra dimensions that form
a fuzzy sphere. The model can then be interpreted as 6-dimensional
gauge theory, with gauge group and geometry depending on the
parameters in the original Lagrangian. We explicitly calculate the tower
of massive Kaluza--Klein modes, consistent with an interpretation as
compactif\/ied higher-dimensional gauge theory, and determine the
ef\/fective compactif\/ied gauge theory. This model has a unique vacuum,
with associated geometry and low-energy gauge group depending only
on the parameters of the potential.

There are many remarkable aspects of this model.  First, it provides
an extremely simple and geometrical mechanism of dynamically
generating extra dimensions,
since it is based on a~basic lesson
from non-commutative gauge theory, namely that non-commutative or
fuzzy spaces can be obtained as solutions of matrix models. The
mechanism is quite generic, and does not require f\/ine-tuning or
supersymmetry. This provides in particular a realization of the
basic ideas of compactif\/ication and dimensional reduction within the
framework of renormalizable quantum f\/ield theory. Moreover, we are
essentially considering a large~$N$ gauge theory, which should allow
to apply the analytical techniques developed in this context.

In particular, it turns out that the generic low-energy gauge group
is given by $SU(n_1) \times SU(n_2)\times U(1)$ or $SU(n)$, while
 gauge groups which are
products of more than two simple components (apart from $U(1)$) do
not seem to occur in this model. The values of $n_1$ and $n_2$ are
determined dynamically. Moreover, a magnetic f\/lux is induced in the
vacua with non-simple gauge group.

The inclusion of fermions in the above class of models showed that the best one could achieve so far is to obtain theories with mirror fermions in bifundamental representations of the low-energy gauge group \cite{Steinacker:2007ay,Chatzistavrakidis:2009ix}. Indeed, studying in detail the fermionic sector of a model which dynamically develops extra dimensions with the geometry of fuzzy $S^2\times S^2$ we found out that the low-energy theory contains two mirror sectors, even when magnetic f\/luxes are included on the two fuzzy spheres. Although mirror fermions do not exclude the possibility to make contact with phenomenology~\cite{Maalampi:1988va}, it would be desirable to obtain exactly chiral fermions.

In order to pursue further the possibility to obtain chiral fermions we introduced an additional structure in the above context, based on orbifolds, in order to obtain chiral low-energy models. In particular we performed a $\Z_3$ orbifold projection of a ${\cal N}=4$ $SU(3N)$ SYM theory, which leads to a ${\cal N}=1$ supersymmetric theory with gauge group $SU(N)^3$. Adding a suitable set of soft supersymmetry breaking terms in the ${\cal N}=1$ theory, certain vacua of the theory were revealed, where twisted fuzzy spheres are dynamically generated. It is well known that the introduction of a soft supersymmetry breaking sector is not only natural but also necessary in the constructions of phenomenologically viable supersymmetric theories, with prime example the case of the MSSM~\cite{Djouadi:2005gj}. Such vacua correspond to models which behave at intermediate energy scales as higher-dimensional theories with a f\/inite Kaluza--Klein tower of massive modes and a~chiral low-energy spectrum. The most interesting chiral models for low-energy phenomenology which can be constructed in this context turn out to be $SU(4)\times SU(2)\times SU(2)$, $SU(4)^3$ and~$SU(3)^3$.

Subsequently, the possibility to achieve further breaking of the above
models down to the MSSM and $SU(3)_c\times U(1)_{\rm em}$ using twisted fuzzy
spheres was studied and it was shown that this is indeed
possible. Thus the spontaneous symmetry breaking of these unif\/ication
groups acquires an interesting geometrical explanation in terms of
twisted fuzzy spheres.
The most interesting case is the trinif\/ication group~$SU(3)^3$,
which can be promoted even to an all-loop f\/inite theory
(for a review see~\cite{Heinemeyer:2010xt}) and therefore
it is suitable to make predictions~\cite{Ma:2004mi,Heinemeyer:2009zs}.

We have thus shown that fuzzy extra dimensions can arise in simple
f\/ield-theoretical models which are chiral, renormalizable,
and may be phenomenologically viable. Moreover,
since some of these models can be
f\/inite with fermions in the adjoint of an underlying~$SU(3N)$ gauge group, these models can be
generalized into the framework of Yang--Mills matrix model
such as~\cite{Ishibashi:1996xs,Aoki:2002jt}.

We have already argued about the importance of the soft supersymmetry breaking terms in the above models. It would be very interesting to explore further the possibility to obtain these terms directly from a higher-dimensional gauge theory. In fact it is known that the SSB terms arise naturally in four dimensions when the ordinary CSDR scheme is used~\cite{Manousselis:2000aj,Chatzistavrakidis:2008ii}. Therefore it is natural to expect that applying
the fuzzy CSDR scheme to an orbifolded higher-dimensional gauge theory coupled to fermions one could naturally obtain the models we discussed above without having to add the SSB terms by hand. We plan to report on this soon.

\appendix

\section{Clif\/ford algebra conventions}\label{appendixA}

In this appendix we collect our conventions on the representations of the Clif\/ford
algebras we have used in Section~\ref{section3}.

The gamma matrices $\gamma_{\mu}$ def\/ine the four-dimensional Clif\/ford algebra and
they are chosen to be purely imaginary, corresponding to the Majorana representation
in four dimensions. In our conventions this representation is
explicitly given by the matrices
\[ \g_0 = \s_0\otimes\s_2,\qquad
       \g_1 = i\s_0\otimes\s_3,\qquad
       \g_2 = i\s_1\otimes\s_1,\qquad
       \g_3 = i\s_3\otimes\s_1,
       \]
where $\s_0:=\one_2$ is the identity matrix.

Moreover, we give here the explicit Majorana representation of the
six-dimensional Clif\/ford algebra, which is known to exist in six Euclidean dimensions. This is naturally adapted to
$SO(3)_L \times SO(3)_R \subset SO(6)$, and closely related to other
constructions, see for example~\cite{Behr:2005wp}. First we consider the
 matrices
\begin{alignat*}{4}
&\gamma^{1}_{L}=\sigma ^{1}\otimes \sigma ^{2}, \qquad &&\gamma^{2}_{L}=\sigma ^{2}\otimes \one,\qquad
&& \gamma^{3}_{L}=\sigma^{3}\otimes \sigma ^{2},& \\
&\gamma^{1}_{R}=\sigma ^{2}\otimes \sigma ^{1},\qquad &&\gamma^{2}_{R}=-\sigma ^{2}\otimes \sigma ^{3},\qquad
&&\gamma^{3}_{R}= \one\otimes \sigma ^{2},*
\end{alignat*}
which are antisymmetric and purely imaginary,
hence hermitian, and they satisfy
\begin{gather*}
\gamma _{L}^{i}\gamma _{L}^{j}   =   \delta ^{ij}+i\epsilon ^{ij}_{k}\gamma _{L}^{k},\qquad
\gamma _{R}^{i}\gamma ^{j}_{R}   =   \delta ^{ij}+i\epsilon ^{ij}_{k}\gamma ^{k}_{R},\qquad
{}[\gamma ^{i}_{L},\gamma ^{j}_{R}]   =   0.
\end{gather*}
Then the following matrices def\/ine a representation of the $SO(6)$ Clif\/ford
algebra
\[
\D_i = i \sigma_1\otimes \gamma_{L}^{i},
\qquad \D_{3+i} = i \sigma_3\otimes \gamma_{R}^{i},
\]
satisfying the desired relation
\[
\{\D^{\mu },\D^{\nu }\} = - 2\delta^{\mu \nu }.
\]
They are manifestly anti-symmetric and real, hence they furnish a Majorana representation.
The left and right chiral projections are given by
\[
\Gamma_L^{(Y)}  =  \D_1\D_2\D_3 =  \sigma_1 \otimes \one,\qquad
 \Gamma_R^{(Y)}  =  \D_4\D_5\D_6 =  \sigma_3 \otimes \one
\]
and the six-dimensional chirality operator is
\[
\Gamma^{(Y)} = -i\Gamma_L^{(Y)}\Gamma_R^{(Y)}
= -\sigma_2 \otimes \one.
\]

\subsection*{Acknowledgements}

 The authors would like to thank P.~Aschieri, T.~Grammatikopoulos, J.~Madore, P.~Manousselis and H.~Steinacker for their collaboration on the subjects of this report over the last years. This work is supported by the NTUA programmes for basic research PEVE 2008 and 2009, the European Union's RTN programme under contract MRTN-CT-2006-035505 and the European Union's ITN programme ``UNILHC'' PITN-GA-2009-237920.

\LastPageEnding

\end{document}